\documentclass[aip,graphicx,preprint,amsmath,amssymb]{revtex4-1}
\usepackage[utf8]{inputenc}
\usepackage{amsfonts}
\usepackage{graphicx}
\usepackage{epstopdf}
\usepackage{hyperref}
\usepackage{courier}
\newcommand{\br}{{\bf r}}
\newcommand{\bn}{{\bf n}}
\newcommand{\bg}{{\bf g}}
\newcommand{\bu}{{\bf u}}

\newcommand{\bq}{{\bf q}}
\begin{document}
\title{Trapping of swimmers in a vortex lattice}
\author{S.~A. Berman}
\email{sberman4@ucmerced.edu}
\affiliation{Department of Physics, University of California, Merced, CA 95344 USA}
\author{K.~A. Mitchell}
\email{kmitchell@ucmerced.edu}
\affiliation{Department of Physics, University of California, Merced, CA 95344 USA}

\date{\today}
\begin{abstract}
We examine the motion of rigid, ellipsoidal swimmers subjected to a steady vortex flow in two dimensions. Numerical simulations of swimmers in a spatially periodic array of vortices reveal a range of possible behaviors, including trapping inside a single vortex and motility-induced diffusion across many vortices. While the trapping probability vanishes at a sufficiently high swimming speed, we find that it exhibits surprisingly large oscillations as this critical swimming speed is approached. Strikingly, at even higher swimming speeds, we find swimmers that swim perpendicular to their elongation direction can again become trapped. To explain this complex behavior, we investigate the underlying swimmer phase-space geometry. We identify the fixed points and periodic orbits of the swimmer equations of motion that regulate swimmer trapping inside a single vortex cell. For low to intermediate swimming speeds, we find that a stable periodic orbit surrounded by invariant tori forms a transport barrier to swimmers and can trap them inside individual vortices. For swimming speeds approaching the maximum fluid speed, we find instead that perpendicular swimmers can be trapped by asymptotically stable fixed points. A bifurcation analysis of the stable periodic orbit and the fixed points explains the complex and non-monotonic breakdown and reemergence of  swimmer trapping as the swimmer speed and shape are varied.
\end{abstract}
\maketitle

\begin{quotation}
The motion of self-propelled agents in complex environments arises in a variety of natural and engineered systems.
Examples include the navigation of aquatic vessels \cite{Rhoads2013a}, the flocking of flying and swimming organisms \cite{Vicsek1995}, and the motion of motile bacteria \cite{Rabani2013,Rusconi2014,Ariel2015} and artificial microswimmers \cite{Bechinger2016,Wilson2012} in fluids.
A great deal of attention has been paid to the consequences of the interaction of many active particles, for example by coupling to the surrounding fluid \cite{Theillard2017,Miles2019} or through direct inter-particle attraction \cite{Redner2013} or alignment. \cite{Vicsek1995}
In these cases, one is typically interested in determining the activity-driven collective flow of particles that emerges despite an initially quiescent state with no macroscopic flow.
However, a macroscopic flow may instead be driven externally, for example using a microfluidic device. \cite{Yazdi2012,Rusconi2014,Sokolov2016}
Whatever the origin of the flow driving individual particles, they ultimately follow trajectories determined by the superposition of their self-propulsion and the macroscopic flow.
If the particles were passive, then it is well-known that their motion would be restricted in space by transport barriers which can be computed from the macroscopic flow field using dynamical systems theory. \cite{Ottino1990,Aref2017}
Comparatively little is known about the analogues of such transport barriers for active particles in fluid flows.
The objective of this paper is to investigate the question of transport barriers to self-propelled agents in an externally-imposed flow using dynamical systems theory.
\end{quotation}

\section{Introduction}
Self-propelled particles, henceforth ``swimmers," are advected by fluid flows like passive particles; however, the addition of activity leads to dramatic changes in the transport properties of the swimmers compared to passive particles. \cite{Torney2007,Khurana2011,Khurana2012,Rusconi2014}
Experiments on motile bacteria show that swimmers follow tumbling, rather than straight, trajectories in a laminar channel flow, and they tend to get trapped in high-shear regions of the flow. \cite{Rusconi2014}
In vortex flows, experiments show that elongated swimming bacteria tend to get ejected from vortex centers and aggregate along their boundaries. \cite{Yazdi2012,Sokolov2016}
Simulations of ellipsoidal swimmers in a steady array of counter-rotating vortices confirm this behavior and furthermore predict that this behavior becomes more pronounced as the swimmer's shape becomes more slender. \cite{Torney2007}
When a simple time-dependence is added to this flow, allowing even passive particles to escape their localized domain and migrate throughout the fluid via chaotic advection, calculations of swimmer transport statistics show a non-monotonic dependence on the swimming speed and swimmer shape. \cite{Khurana2012}
Counter-intuitively, this is due to the hydrodynamic trapping of swimmers in certain regions of the flow for times significantly exceeding the trapping times of passive particles in the same flow. \cite{Khurana2011}

Some progress has been made in uncovering the swimmer phase-space geometry underlying these phenomena.
For a rigid, non-deformable swimmer, the phase space consists of all possible positions and orientations of the swimmer, in contrast to the passive particle phase space, where the particle's position provides the only relevant degrees of freedom for transport.
There are several studies on the structure of swimmer phase space for laminar flows with continuous symmetry.
It has been shown that the tumbling trajectories of swimmers in translationally-invariant laminar channel (or pipe) flow are due to continuous families of periodic or quasi-periodic orbits (i.e. invariant tori) that foliate the swimmer's phase space. \cite{Zottl2012,Zottl2013}
Meanwhile, the motility-induced trapping of swimming bacteria in high-shear regions of the flow is due to the interplay between this phase-space structure and noise in the swimmer's orientation. \cite{Rusconi2014}
The case of rotationally-invariant two-dimensional (2D) vortex flows is similar, where an analysis in a rotating frame reveals that ellipsoidal swimmers can get trapped in a region of phase space consisting of a continuous family of periodic orbits surrounding a stable fixed point. \cite{Torney2007,Arguedas-Leiva2019}
In these cases, the continuous symmetry results in the foliation of phase space by regular, non-chaotic trajectories.
However, for generic laminar flows, one expects to observe chaos, or possibly a mixed phase space containing both regular and chaotic regions, due to the fact that the swimmer phase space is at least three dimensional (3D).
Indeed, a recent work \cite{Ariel2019} attributes the L\'{e}vy statistics of diffusing swimmers in a periodic array of vortices \cite{Ariel2017} to the underlying mixed phase space.
In the case of swimmer motion in viscoelastic fluids, simulations suggest that the swimmer phase space can also contain attracting limit cycles, which may trap swimmers in vortices. \cite{Ardekani2012}

In this paper, we study transport barriers to active particles in a steady 2D flow without continuous symmetry.
Specifically, we elucidate the phase-space structures that lead to the trapping of rigid ellipsoidal swimmers in individual vortices of the steady, spatially periodic vortex flow.  \cite{Torney2007,Khurana2011,Khurana2012}
Our investigation is framed around understanding the probability of trapping as three key parameters are varied: the swimmer's speed, shape, and relative swimming direction, taken to be either parallel or perpendicular to the major axis of the ellipse.
We report surprisingly large oscillations of the trapping probability as the swimming speed is increased, particularly for elongated swimmers swimming perpendicular to their major axes.
Strikingly, for these perpendicular swimmers, we also find that trapping ceases for a range of intermediate swimming speeds and then reemerges at relatively high swimming speeds.
We show that trapping occurs due to the existence of stable periodic orbits surrounded by invariant tori or due to attracting fixed points or limit cycles, and we identify the bifurcations that lead to the breakdown of these transport barriers as the swimmer parameters are varied.
We find both local and global bifurcations, involving nearby fixed points, and we show that these bifurcations provide insight into the non-monotonic dependence of the trapping probability on swimming speed.
Lastly, we demonstrate numerically that the destruction of the transport barriers accurately predicts the parameters where the trapping probability vanishes.

This paper is organized as follows.
In Sec.~\ref{sec:model}, we introduce the model equations for the swimmer dynamics and the fluid flow, and discuss the relevant symmetries of the equations.
In Sec.~\ref{sec:numerics}, we present numerical calculations of the probability of swimmer trapping inside a vortex as swimmer speed, shape, and relative swimming direction are varied.
In Sec.~\ref{sec:eq}, we investigate the fixed points of the swimmer equations of motion and show how changes in their linear stability properties explain the onset of trapping of perpendicular swimmers with high swimming speed.
In Sec.~\ref{sec:po}, we numerically compute the periodic orbits responsible for trapping up to intermediate swimming speeds for the full range of swimmer shapes and relative swimming directions, and we show how bifurcations of these orbits can be used to understand the complex dependence of the trapping probability on the swimmer parameters.
Finally, in Sec.~\ref{sec:concl}, we make concluding remarks.

\section{ODE model of swimmer dynamics}\label{sec:model}
\begin{figure}
\centering
\includegraphics[width=0.9\textwidth]{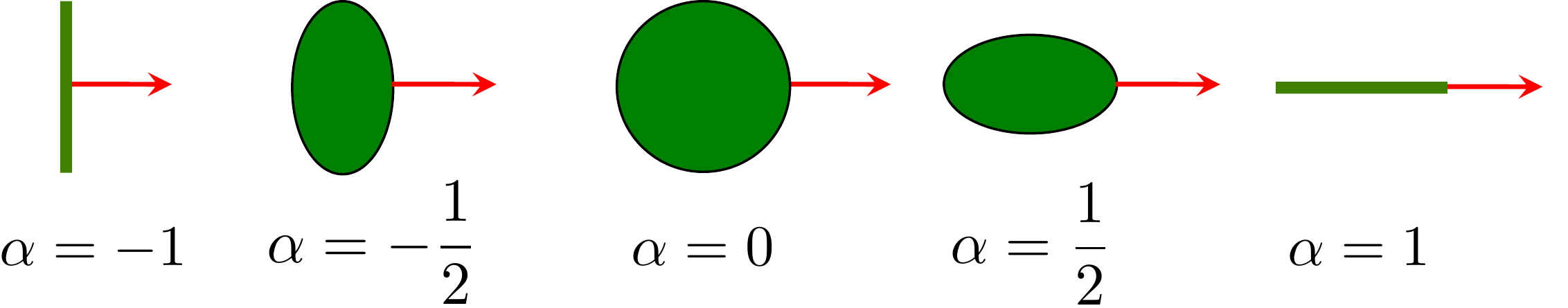}
\caption{Schematic illustrating the particle shapes and relative swimming directions (red arrows) as $\alpha$ is varied between $-1$ and $1$.}\label{fig:alpha_cartoon}
\end{figure}
We consider the motion of an ellipsoidal particle in two dimensions, with position $\br = (x,y)$ and orientation $\hat{\bn} = (\cos\theta, \sin \theta)$ that swims in a steady, i.e.\ time-independent, fluid flow with velocity $\bu(\br)$.
We shall denote the phase-space coordinates of the swimmer as $\bq = (\br,\theta)$.
Assuming the size of the particle is small compared to the length scale over which $\bu$ varies appreciably, the equations of motion for the swimmer are \cite{Torney2007}
\begin{subequations}\label{eq:model_general}
\begin{align}
\dot{\br} & = \bu + v_0 \hat{\bn}, \\
\dot{\theta} & = \frac{\omega_z}{2}  +\alpha\, \hat{\bg} \cdot {\bf E} \hat{\bn},
\end{align}
\end{subequations}
where $\omega_z = \hat{\bf z} \cdot (\nabla \times \bu)$ is the $z$-component of the vorticity, $\hat{\bg} = (-\sin \theta,\cos \theta)$ is a unit vector perpendicular to $\hat{\bn}$ and ${\bf E} = (\nabla \bu + \nabla \bu^{\rm T})/2$  is the symmetric rate-of-strain tensor.
The parameter $\alpha$ characterizes the shape of the ellipse and the relative swimming direction, with $-1 \leq \alpha \leq 1$.
It is defined as $|\alpha| = (1-\gamma^2)/(1+\gamma^2)$, where $\gamma$ is the ratio of the minor to major axes of the ellipse.
As illustrated in Fig.~\ref{fig:alpha_cartoon}, negative values of $\alpha$ correspond to swimming perpendicular to the major axis of the ellipse, while positive values correspond to swimming along the major axis.
The limiting case $|\alpha| = 1$ corresponds to an infinitely slender rod shape, while $\alpha = 0$ is a circle.
The particular case $\alpha = -1$ coincides with the equations of motion for an infinitesimal line element of a front propagating with constant speed $v_0$ in the local fluid frame, which applies to chemical reaction fronts \cite{Mahoney2012,Mitchell2012} and reachability fronts \cite{Rhoads2013a} in externally driven fluid flows.
We shall refer to $\alpha < 0$ ($\alpha > 0$) swimmers as perpendicular (parallel) swimmers.

There is a correspondence between the 2D particle shapes in our model and 3D axisymmetric particles with the swimming direction as the symmetry axis. \cite{Borgnino2019}
Specifically, $\alpha < 0$ corresponds to oblate spheroids, $\alpha=0$ to spheres, and $\alpha > 0$ to prolate spheroids.
Self-propelled prolate spheroids and spheres have been the focus of most theoretical and modeling studies, due to the profusion of biological and artificial microswimmers with those shapes, including bacteria and Janus particles. \cite{Bechinger2016}
Recently, however, increasing attention has been paid to self-propelled oblate spheroids. \cite{Borgnino2019,Arguedas-Leiva2019}
Experimentally, self-propelled oblate spheroids have been created in the form of vesicles fueled by chemical reactions catalyzed by platinum nanoparticles,\cite{Wilson2012} and they may one day be realized as droplets coated with an active nematic. \cite{Tjhung2012,Sanchez2012,Giomi2014}

\subsection{Vortex lattice}
As a model for a steady flow without continuous symmetry, we take $\bu= \left(\psi_{,y},-\psi_{,x}\right)$ with stream function $\psi(\br) = \sin(2\pi x) \sin(2\pi y) /2\pi$, which corresponds to a vortex lattice: a spatially periodic array of alternating vortices.
This serves as a simplified model for thermally convective \cite{Solomon1988}, magnetohydrodynamically driven \cite{Solomon2003,Mahoney2015}, and surface-wave driven \cite{Francois2017} fluid flows realized in experiments, which may also have a controllable time-dependence in the form of a lateral oscillation of the vortex centers.
Due to its widespread study in the transport of passive tracers, this model has also received attention in the transport of self-propelling entities, including chemical reaction fronts,\cite{Abel2001,Cencini2003,Mahoney2012,Mitchell2012,Xin2013} autonomous underwater vehicles, \cite{Rhoads2013a} and microswimmers. \cite{Torney2007,Khurana2011,Ariel2017}
For the vortex lattice, the equations of motion \eqref{eq:model_general} become
\begin{subequations}\label{eq:model_vortex}
\begin{align} \label{eq:xdot}
\dot{x} & = \sin (2 \pi x) \cos (2 \pi y) + v_0 \cos \theta, \\ \label{eq:ydot}
\dot{y} & = -\cos (2\pi x) \sin (2\pi y)+ v_0 \sin \theta, \\ \label{eq:thdot}
\dot{\theta} & = 2 \pi \left[ \sin (2\pi x) \sin (2\pi y) - \alpha \cos(2\pi x) \cos( 2 \pi y) \sin (2\theta) \right].
\end{align}
\end{subequations}
Because $|\bu| \leq 1$ everywhere, $v_0$ can be thought of as the ratio of the swimming speed to the maximum fluid speed.
In this paper, we restrict our attention to $v_0 \leq 1$, so that there is always some part of space where the swimming speed does not exceed the fluid speed.

\begin{figure}
\centering
\includegraphics[width=0.7\textwidth]{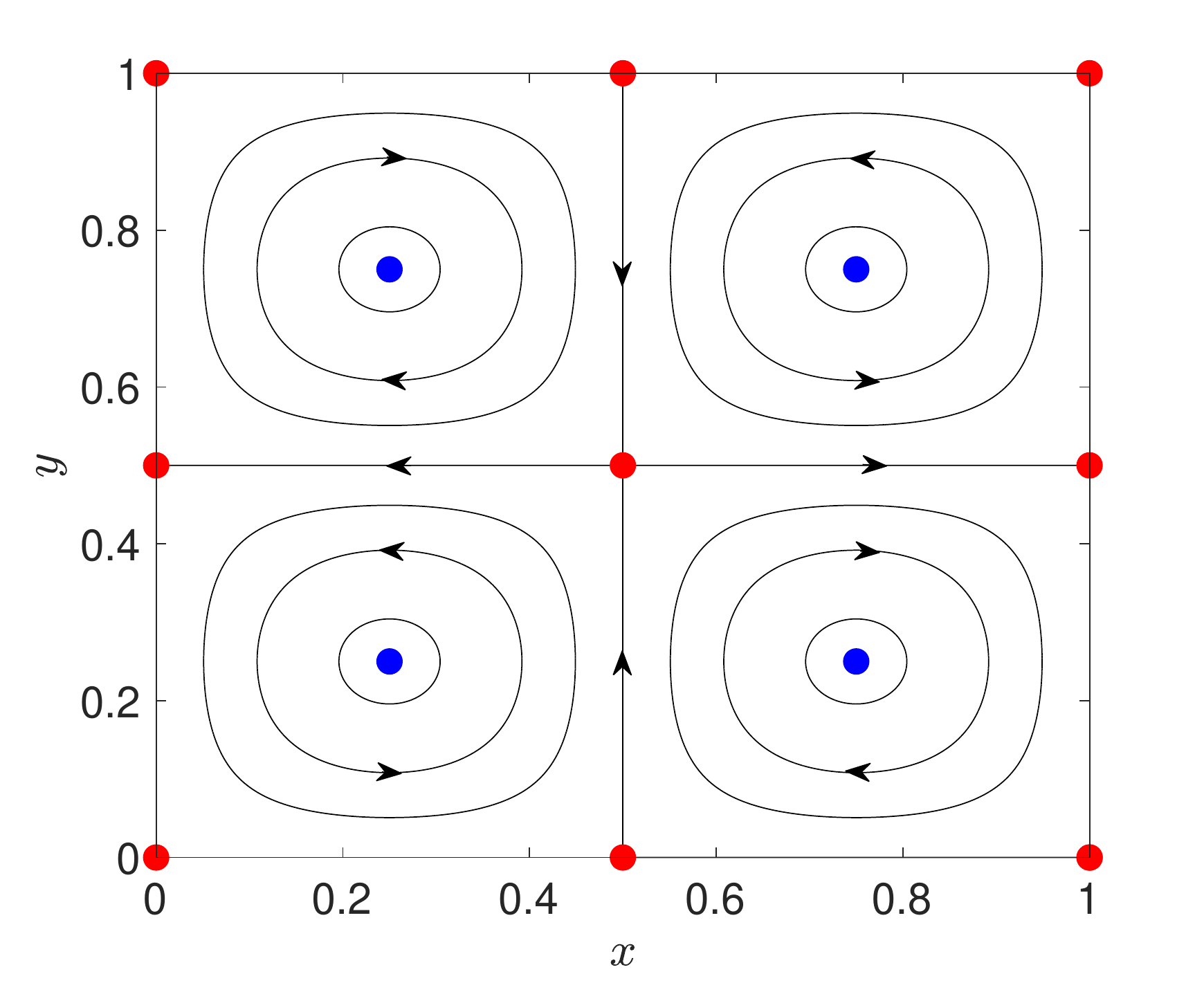}
\caption{Phase portrait for passive tracers in the spatially periodic vortex lattice flow. The blue dots are stable, elliptic fixed points, while the red dots are unstable, hyperbolic fixed points. The black curves are periodic orbits, coinciding with the level curves of the stream function $\psi(x,y)$.}\label{fig:passive_phase_portrait}
\end{figure}
Figure \ref{fig:passive_phase_portrait} shows the phase-portrait for the passive tracer case, i.e.\  Eqs.~\eqref{eq:xdot} and \eqref{eq:ydot} with $v_0=0$, for one spatial period in the $x$- and $y$-directions.
Fluid elements move in square, counter-rotating vortex cells, with the edges of each square being impenetrable separatrices.
At the center of each vortex cell is a stable, elliptic fixed point, i.e.\ a fixed point with purely imaginary eigenvalues.
Surrounding each elliptic fixed point is a continuous family of periodic orbits, each lying on a particular level curve of the stream function $\psi$.
At the corner of each square vortex cell is an unstable, hyperbolic fixed point, i.e. a fixed point with one positive and one negative eigenvalue.
Each hyperbolic fixed point is connected to its neighboring hyperbolic fixed points by heteroclinic orbits.
These orbits are in fact the separatrices sealing off each vortex cell, preventing transport of fluid particles across the cell boundaries (in the absence of molecular diffusion).
In the following, we shall see that each of the aforementioned phase-space structures has an analogue for active particles, i.e.\ when $v_0 \neq 0$ and Eq.~\eqref{eq:thdot} is taken into account. 
However, going from the 2D phase space of passive tracers to the 3D phase space of swimmers drastically changes the geometry of the phase-space structures, which has profound consequences on the transport properties of swimmers in the vortex lattice.

\subsection{Symmetries}
Equations \eqref{eq:model_vortex} possess certain symmetries that are key to understanding the phase space structure.
We first describe the transformations of the phase space coordinates $(x,y,\theta)$ that leave the equations of motion unchanged.
Equations \eqref{eq:model_vortex} are invariant with respect to shifts in $x$ and $y$: $(x,y,\theta) \mapsto (x+1,y,\theta)$, $(x,y,\theta) \mapsto (x,y+1,\theta)$ and $(x,y,\theta) \mapsto (x+1/2,y+1/2,\theta)$.
They are also invariant with respect to rotations about any given vortex cell center by $\pi/2$.
For example, shifting the origin of coordinates to the center of the lower left vortex in Fig.~\ref{fig:passive_phase_portrait}, rotating both the coordinates and the swimmer orientation by $\pi/2$, and shifting the origin back to its original position gives the transformation $(x,y,\theta) \mapsto (-y + 1/2, x, \theta + \pi/2)$, and this transformation leaves the equations of motion unchanged.

\begin{table}
\centering
\begin{tabular}{r | c }
$t$-symmetry & $(x,y,\theta) \mapsto $ \\ \hline
reflection about vertical & $(-x+\frac{1}{2},y,-\theta)$ \\
reflection about horizontal & $(x,-y+\frac{1}{2},\pi-\theta)$ \\
reflection about $y=x$ & $(y,x, \frac{3\pi}{2} - \theta)$ \\
reflection about $y=-x$ & $(-y,-x,\frac{\pi}{2}-\theta)$ 
\end{tabular}
\caption{Summary of the relevant reversing symmetries, or $t$-symmetries, of a swimmer in a vortex lattice [Eqs.~\eqref{eq:model_vortex}]. Here ``horizontal" and ``vertical" refer to axes going through the vortex center at $\br = (1/4,1/4)$.}\label{tab:tsymm}
\end{table}
In addition, Eqs.~\eqref{eq:model_vortex} are \emph{reversible}, meaning the equations of motion are invariant when combining a specific type of phase-space symmetry, called an involution, with time reversal, i.e.\ sending $t \mapsto -t$. \cite{Roberts1992,Lamb1998}
An involution is simply a symmetry operation which is its own inverse, meaning applying it to the phase space coordinates twice is equivalent to the identity operation, or not changing the coordinates at all.
Thus, reversibility intuitively means that integrating a given initial condition forward in time and ``flipping" it is equivalent to ``flipping" the initial condition and integrating backward in time.
The combination of involution and time reversal is referred to as a reversing symmetry, or $t$-symmetry.
In fact, Eqs.~\eqref{eq:model_vortex} are multiply reversible, meaning there is more than one distinct $t$-symmetry.
The most relevant $t$-symmetries to this article are summarized in Table \ref{tab:tsymm}.
In this case, the $t$-symmetries consist of a reflection $R$ about some axis, where $\br \mapsto R \br$ and the new $\theta$ is chosen so that the swimmer orientation changes as $\hat{\bn} \mapsto -R \hat{\bn}$.
For example, one $t$-symmetry is given by reflection about a vertical axis going through the center of a vortex cell, e.g.\ $(x,y,\theta) \mapsto (-x+1/2,y,-\theta)$, which leaves the equations of motion unchanged upon time reversal.
Due to the $\pi/2$ rotational symmetry, reflection about a horizontal axis going through a vortex center,  e.g.\  $(x,y,\theta) \mapsto (x,-y+1/2,\pi-\theta)$, is also a $t$-symmetry.
In Table \ref{tab:tsymm} we give the transformations of some of the additional $t$-symmetries associated with reflection about an axis going through the diagonal of a vortex cell square, such as the $y=x$  and $y=-x$ axes.
The reversibility of Eqs.~\eqref{eq:model_vortex} implies that the phase space can exhibit a mixture of conservative and dissipative behavior.\cite{Roberts1992,Politi86}
In other words, the system may have a mixed phase space consisting of elliptic islands and chaotic seas, as in the case of Hamiltonian systems, but it may also have attractors and repellers.

Throughout the paper, we will consider both the continuous-time flow generated by Eqs.~\eqref{eq:model_vortex} and the discrete-time map obtained from a Poincar\'{e} surface of section, i.e.\ the Poincar\'{e} map.
The reversibility of Eqs.~\eqref{eq:model_vortex} imparts reversibility to the Poincar\'{e} map, where the discrete-time equivalent of time-reversal is taking the inverse of the map.
We shall consider surfaces of section defined by a fixed value of $\theta$, i.e.\ $\theta = {\rm const} \,({\rm mod}\,2\pi)$.
For a choice of $\theta$ that is invariant under one of the $t$-symmetries shown in Table \ref{tab:tsymm}, it is straightforward to obtain the corresponding $t$-symmetry of the resulting Poincar\'{e} map.
Specifically, in this paper we consider at various points the sections defined by $\theta = 0$ and  $\theta = -\pi/4$.
Because $\theta = 0$ is unchanged by the $t$-symmetry ``reflection about vertical" from Table \ref{tab:tsymm}, then the corresponding $t$-symmetry of the Poincar\'{e} map is simply reflection about the $x = 1/4$ axis, i.e.\ $(x,y) \mapsto (-x + 1/2,y)$.
On the other hand, $\theta = -\pi/4$ is invariant under ``reflection about $y=x$;" hence, the corresponding $t$-symmetry of the Poincar\'{e} map is reflection about the $y=x$ axis, given by $(x,y) \mapsto (y,x)$.

\section{Probability and phase-space geometry of trapping}\label{sec:numerics}
Like passive particles, swimmers that are initially located inside a vortex cell may remain trapped in that cell indefinitely.
Unlike passive particles, however, trapping is not guaranteed.
Rather, it is determined by the swimmer's initial conditions $\bq_0$ and the parameters $v_0$ and $\alpha$.\cite{Torney2007}
Swimmers that are not trapped by definition eventually escape from the vortex cell where they begin and move into neighboring vortex cells, potentially exhibiting long-range transport.\cite{Ariel2019}
For parallel swimmers, i.e.\ $0 \leq \alpha \leq 1$, the breakdown of trapping in the periodic vortex lattice---where all initial conditions result in escape---has been investigated in Ref.~\onlinecite{Torney2007} and has been shown to occur once the swimming speed $v_0$ exceeds a critical value $v_0^*(\alpha)$, with $v_0^*(1) = 0$ and $d v_0^*/d\alpha < 0$ for all $\alpha \geq 0$.
In other words, for this range of $\alpha$, there is a minimum swimming speed $v_0^*$ for guaranteed escape, which decreases as the particle shape becomes more elongated, and vanishes in the limit of infinitely slender rods.
The existence of $v_0^*$ accords with the basic intuition that a particle that swims fast enough should be able to escape a vortex, regardless of its initial conditions.
Does this result hold for perpendicular swimmers, with $\alpha < 0$?
And does this intuition imply further that the probability of trapping with a random initial condition monotonically decreases as $v_0$ increases?
Also, does escape from the initial vortex cell guarantee long-range migration of the swimmer, or do some escaping swimmers remain effectively trapped, i.e.\ localized in the vicinity of the initial vortex cell?

\begin{figure}
\centering
\includegraphics[width=0.9\textwidth]{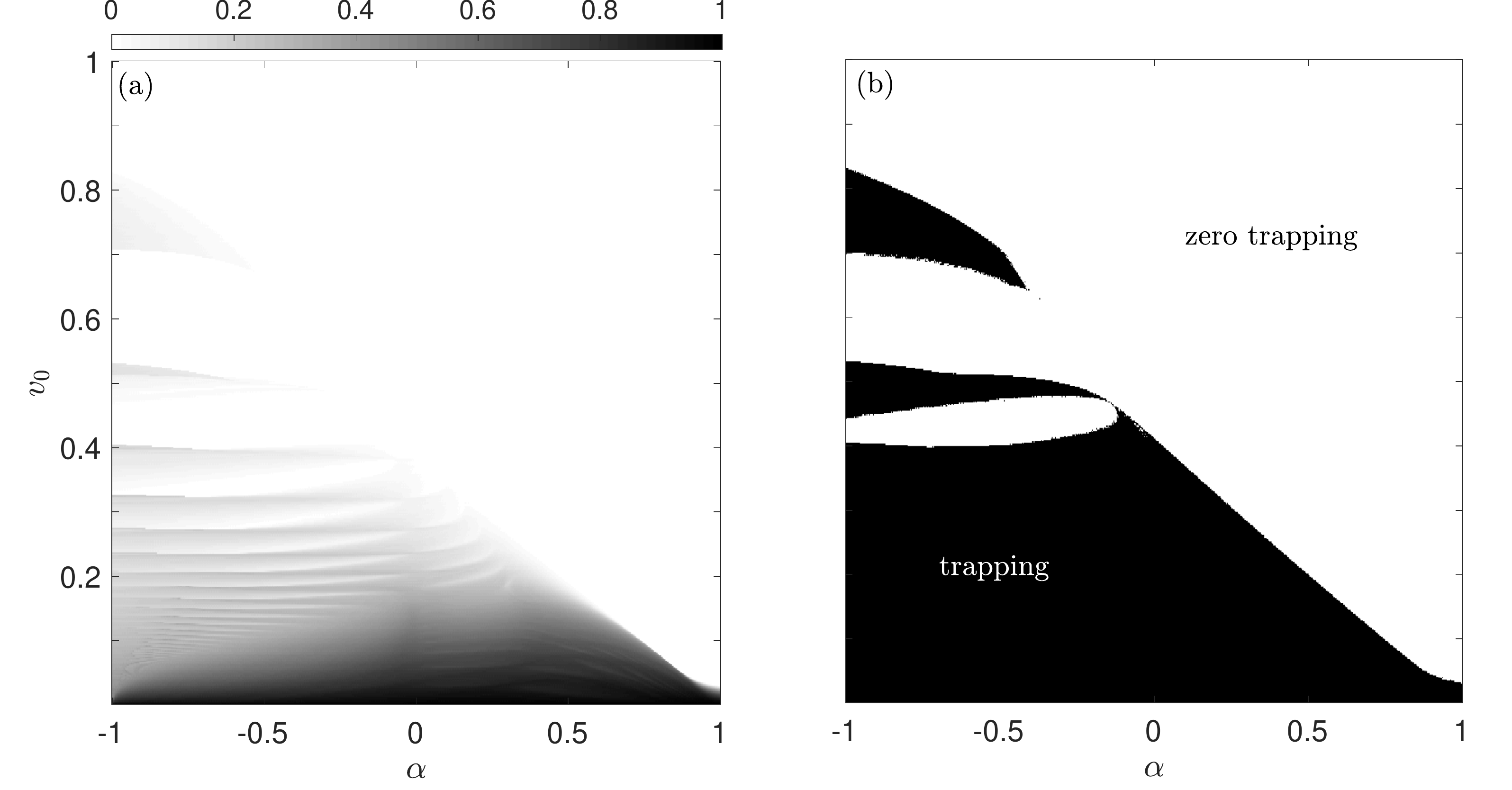}
\caption{Probability $P_{\rm trap}$ of swimmer trapping in an individual vortex cell for times $t \leq 40$ as a function of $v_0$ and $\alpha$.  (a) The probability $P_{\rm trap}$ is indicated by the gray scale, with darker shading corresponding to higher probability. (b) Binary version of (a), where $(v_0,\alpha)$ points with nonzero $P_{\rm trap}$ values are mapped to black and those with $P_{\rm trap}=0$ are mapped to white.}\label{fig:monte_carlo}
\end{figure}

We address these questions by numerically calculating the probability of trapping and localization as a function of $v_0$ and $\alpha$.
For each selected combination of $(v_0,\alpha)$, we numerically integrate Eqs.~\eqref{eq:model_vortex} for $10^4$ initial conditions with positions uniformly randomly distributed inside one vortex cell, i.e. with $(x_0,y_0) \in [0,1/2]\times [0,1/2]$, and uniformly randomly distributed orientations in the full range $\theta_0 \in [0,2\pi]$.
We employ an adaptive Runge-Kutta $(4,5)$ scheme (implemented in Matlab as \texttt{ode45}) and integrate each trajectory from $t=0$ to a final time $t_f = 40$.
The probability of trapping $P_{\rm trap}(v_0,\alpha)$ is defined as the fraction of trajectories that remain within the initial vortex cell $[0,1/2]\times [0,1/2]$ for the entire integration time.
On the other hand, we define the probability of localization $P_{\rm loc}(v_0,\alpha)$ as the fraction of trajectories that remain within the circle of radius $1/4$ circumscribing the initial vortex cell for the entire integration time.
Hence, $P_{\rm loc} \geq P_{\rm trap}$.
We investigate the parameter range $v_0 \in [0.002,1]$ and $\alpha \in [-1,1]$, using small increments $\Delta v_0 = 0.002$ and $\Delta \alpha = 0.002$.
\begin{figure}
\centering
\includegraphics[width=0.9\textwidth]{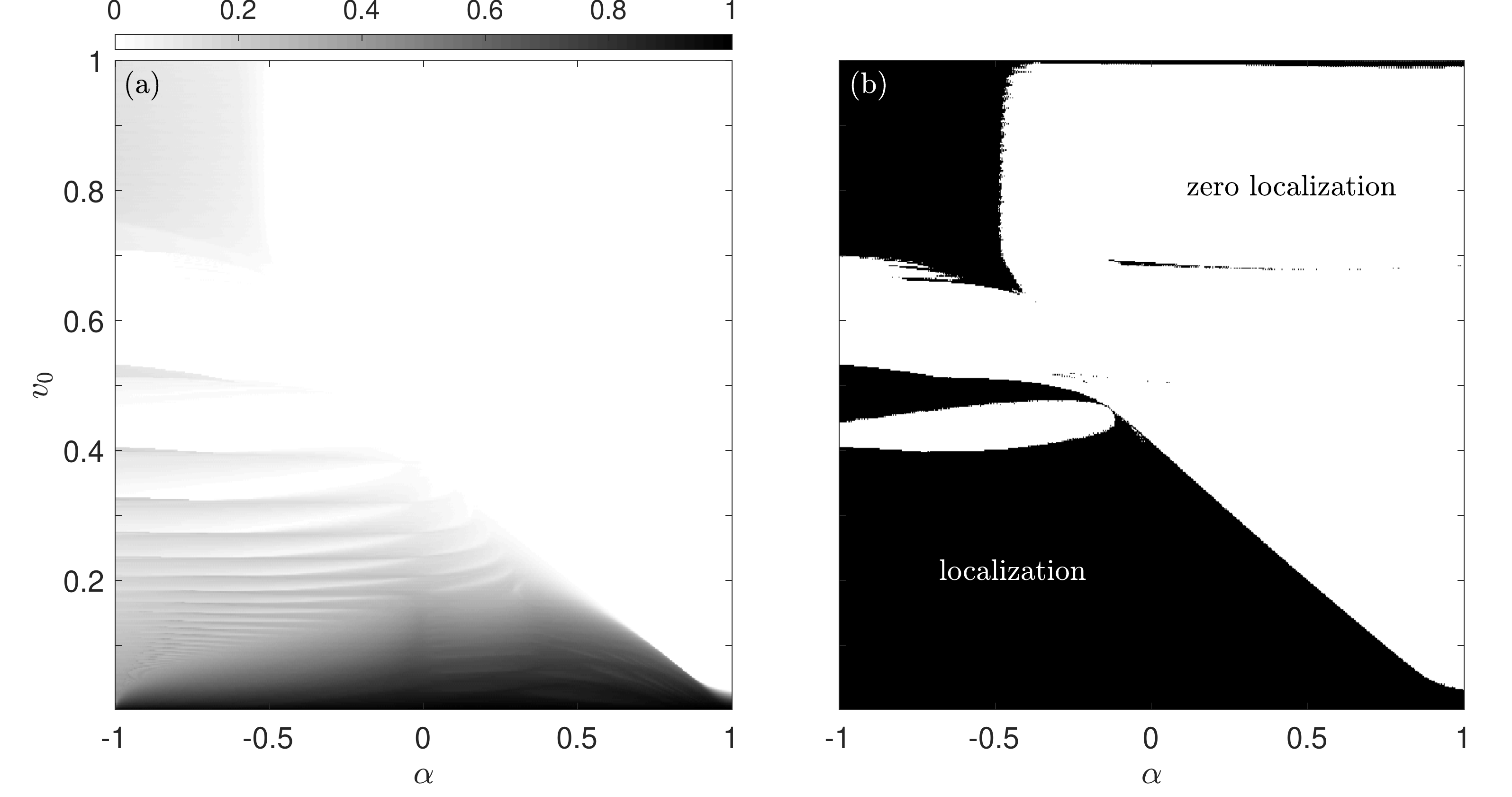}
\caption{Probability $P_{\rm loc}$ of swimmer localization in a circle circumscribing an individual vortex cell for times $t \leq 40$ as a function of $v_0$ and $\alpha$.  (a) The probability $P_{\rm loc}$ is indicated by the gray scale, with darker shading corresponding to higher probability. (b) Binary version of (a), where $(v_0,\alpha)$ points with nonzero $P_{\rm loc}$ values are mapped to black and those with $P_{\rm loc}=0$ are mapped to white.}\label{fig:monte_carlo_loc}
\end{figure}

Figure \ref{fig:monte_carlo} shows the result of the trapping calculation, with the values of $P_{\rm trap}(v_0,\alpha)$ plotted in Fig.~\ref{fig:monte_carlo}a and the distinction between zero and nonzero values of $P_{\rm trap}$ plotted in Fig.~\ref{fig:monte_carlo}b.
As expected, $P_{\rm trap} \rightarrow 1$ as $v_0 \rightarrow 0$, i.e. approaching the limit of passive tracers.
Also, for each $\alpha$ there is clearly a critical swimming speed $v_0^*(\alpha)$ above which all swimmers escape the initial vortex cell, including perpendicular swimmers, i.e. $P_{\rm trap}(v_0,\alpha) = 0$ for all $v_0 > v_0^*(\alpha)$ (Fig.~\ref{fig:monte_carlo}b).
Surprisingly, however, there is a large range of $\alpha < 0$ for which trapping ceases completely as swimming speed is increased, before returning at even higher swimming speeds.
For most perpendicular swimmers, trapping first ceases around $v_0 = 0.4$ (the finger-shaped white region of Fig.~\ref{fig:monte_carlo}b), returns around $v_0 = 0.45$, and then ceases again around $v_0 = 0.5$.
This sequence then repeats at even higher swimming speeds beginning around $v_0 = 0.7$, albeit for a slightly smaller range of $\alpha$ values.
Even though escape is guaranteed when $v_0$ becomes large enough, the breakdown of trapping is non-monotonic with respect to $v_0$ for these swimmers.
Perhaps even more striking is the non-monotonicity of $P_{\rm trap}(v_0,\alpha)$ in the nonzero regions, seen in Fig.~\ref{fig:monte_carlo}a.
Rather than steadily decreasing as $v_0$ increases, $P_{\rm trap}(v_0,\alpha)$ can oscillate widely.
This effect is particularly pronounced for the perpendicular swimmers, where $P_{\rm trap}$ nearly goes to zero repeatedly as $v_0$ increases, but is also present for parallel swimmers.
The oscillations for $v_0 \lesssim 0.4$  resemble a regular pattern that is repeated as  $v_0$ decreases with decreasing spacing in $v_0$.

Meanwhile, Fig.~\ref{fig:monte_carlo_loc} shows the probability $P_{\rm loc}$ of a swimmer remaining localized in a circle surrounding the initial vortex cell for the duration of the calculation.
Comparing Figs.~\ref{fig:monte_carlo_loc} and \ref{fig:monte_carlo}, we see that $P_{\rm loc}$ and $P_{\rm trap}$ appear identical for most values of the parameters.
Thus, we conclude that most of the time, swimmers which escape the initial vortex cell do not remain localized in a region surrounding that cell, or conversely, localization typically implies trapping.
However, there is a significant exception to this in the range of parameters  $v_0 \gtrsim 0.8$ and $\alpha \lesssim -0.5$, where swimmers can be localized with a significant probability $P_{\rm loc}$ (Fig.~\ref{fig:monte_carlo_loc}a) while escaping the initial vortex cell at some point with probability one (Fig.~\ref{fig:monte_carlo}b).
We note that there are other small ranges of parameters where $P_{\rm loc} \neq 0$ and $P_{\rm trap} = 0$ (e.g. $v_0 \approx 0.7$ and $-0.1 \lesssim \alpha \lesssim 0.5$), as seen by comparing Figs.~\ref{fig:monte_carlo_loc}b and \ref{fig:monte_carlo}b, though the fact that these regions are not visible on Fig.~\ref{fig:monte_carlo_loc}a indicates that the probability of these events is quite small.
Hence, we find that escape from a vortex cell does not always guarantee long-range migration for a swimmer: in this case, certain escaping perpendicular swimmers can still be effectively trapped in a region surrounding the initial vortex cell.

\begin{figure}
\includegraphics[width=0.9\textwidth]{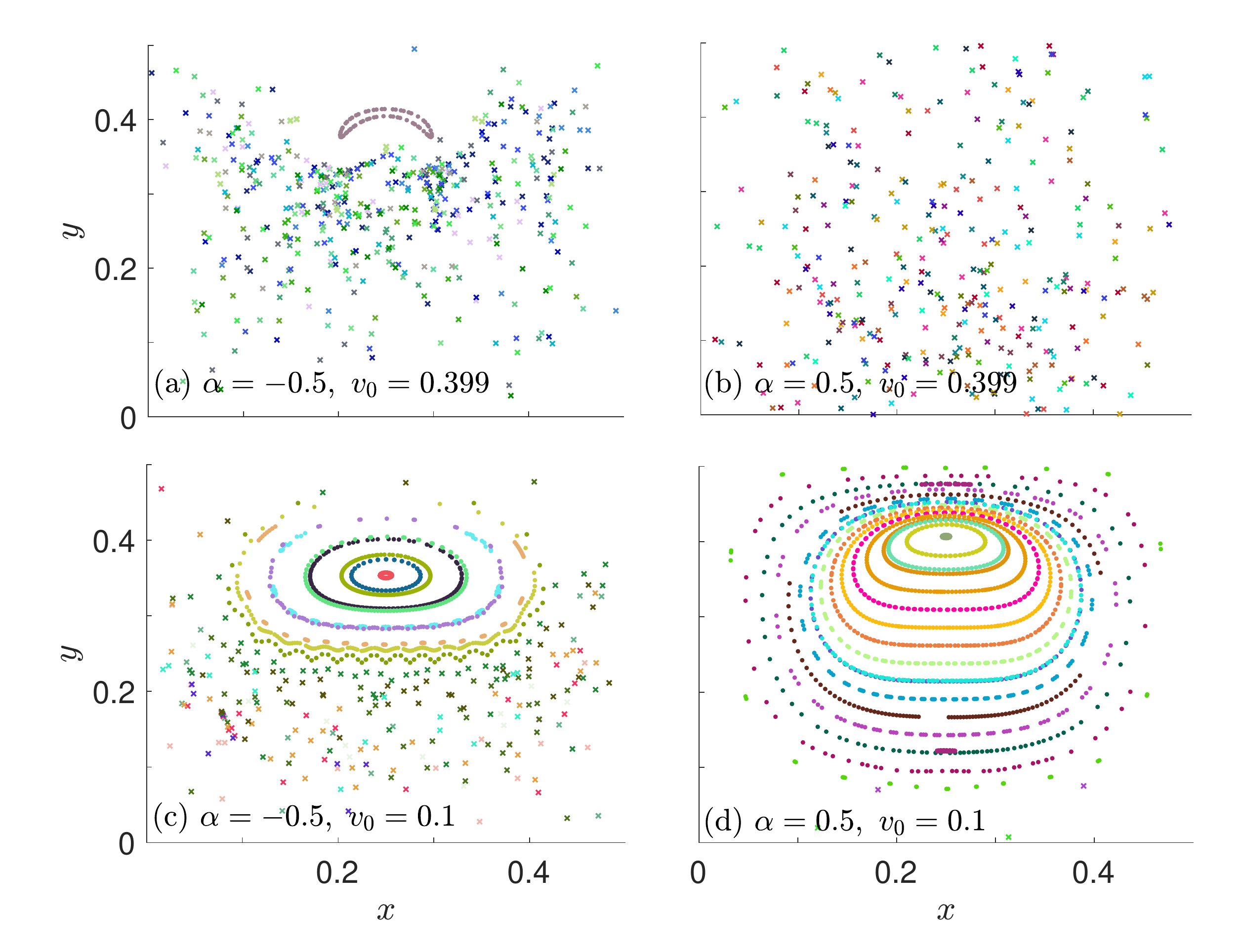}
\caption{Poincar\'{e} section $\theta=0$ with $\dot{\theta} >0$ restricted to one vortex cell, for different values of the parameters. (a) $\alpha = -0.5$, $v_0 = 0.399$. (b) $\alpha = 0.5$, $v_0 = 0.399$. (c) $\alpha = -0.5$, $v_0 = 0.1$. (d) $\alpha = 0.5$, $v_0 = 0.1$. In all figures, different colors correspond to different initial conditions, and dots correspond to swimmers that remain trapped in the initial vortex, while exes correspond to swimmers that escaped the initial vortex at some point in time.}\label{fig:poincare}
\end{figure}
By looking at the swimmer trajectories using a Poincar\'{e} surface of section, we can assess the phase space structures that are responsible for trapping.
We find it convenient to take the surface of section $\theta = 0$, with $\dot{\theta} > 0$.
In Fig.~\ref{fig:poincare}, we show typical Poincar\'{e} sections for different values of $v_0$ and $\alpha$, covering both parallel swimmers (Figs.~\ref{fig:poincare}b and \ref{fig:poincare}d) and perpendicular swimmers (Figs.~\ref{fig:poincare}a and \ref{fig:poincare}c).
The sections are calculated by selecting 20 initial conditions within one vortex cell, evenly spaced on the line $x =  0.25$ with $y \in (0,0.5)$ and $\theta = 0$, and integrated until $t_f = 200$.
We show the  intersections $(x,y) \mod 1$ of the swimmer trajectories with the surface of section, so that we can see swimmers that leave the initial vortex and migrate throughout the lattice.
For low to intermediate values of $v_0$, the typical scenario we observe is that trapped swimmers (dots tracing out closed curves in Fig.~\ref{fig:poincare}) move on invariant tori, i.e.\ quasi-periodic trajectories, while escaping swimmers (scattered exes in Fig.~\ref{fig:poincare}) move in a chaotic sea.
At high enough swimming speeds where we observe no trapping, i.e.\ $P_{\rm trap}(v_0,\alpha) = 0$, the entire phase space is often a chaotic sea (Fig.~\ref{fig:poincare}b).
When the swimming speed is low and probability of trapping is quite high, most of the phase space is filled with invariant tori (Fig.~\ref{fig:poincare}d), and the area occupied by the tori decreases as $P_{\rm trap}$ decreases (Fig.~\ref{fig:poincare}c).
As $P_{\rm trap}$ approaches zero, so too does the area of the tori corresponding to trapped swimmer motion (Fig.~\ref{fig:poincare}a).

Remarkably, the Poincar\'{e} sections for the full range of swimmer shapes and swimming speeds have a very similar structure.
This is particularly evident for the trapped swimmers, for which almost all invariant tori appear to surround a stable periodic orbit on the $t$-symmetry axis $x=1/4$ (Figs.~\ref{fig:poincare}a, \ref{fig:poincare}c, and \ref{fig:poincare}d).
In Sec.~\ref{sec:po}, we show that there is indeed a stable periodic orbit at the center of these tori, and as such it plays a role analogous to the stable fixed point at the vortex center for passive particles.
Specifically, the nested family of tori around the stable periodic orbit traps swimmers inside the vortex cell like the nested family of periodic orbits around the stable fixed point traps passive particles.
On Poincar\'{e} sections for high swimming speeds $v_0 \gtrsim 0.7$ and $\alpha$ such that  $P_{\rm loc} \neq 0$ (which includes the high-$v_0$ region where $P_{\rm trap} \neq 0$), however, we do not observe invariant tori associated with either trapped or localized swimmers (not shown).
This suggests an alternative mechanism for swimmer localization at these parameter values.
In the next section, we investigate the fixed points of Eqs.~\eqref{eq:model_vortex}, and we show that the presence of asymptotically-stable fixed points underlies the high-$v_0$ trapping and localization observed in Figs.~\ref{fig:monte_carlo} and \ref{fig:monte_carlo_loc}.
We note that the fixed points do not lie on the surface of section.

\section{Equilibria and high-$v_0$ trapping}\label{sec:eq}
\begin{figure}
\centering
\includegraphics[width=0.7\textwidth]{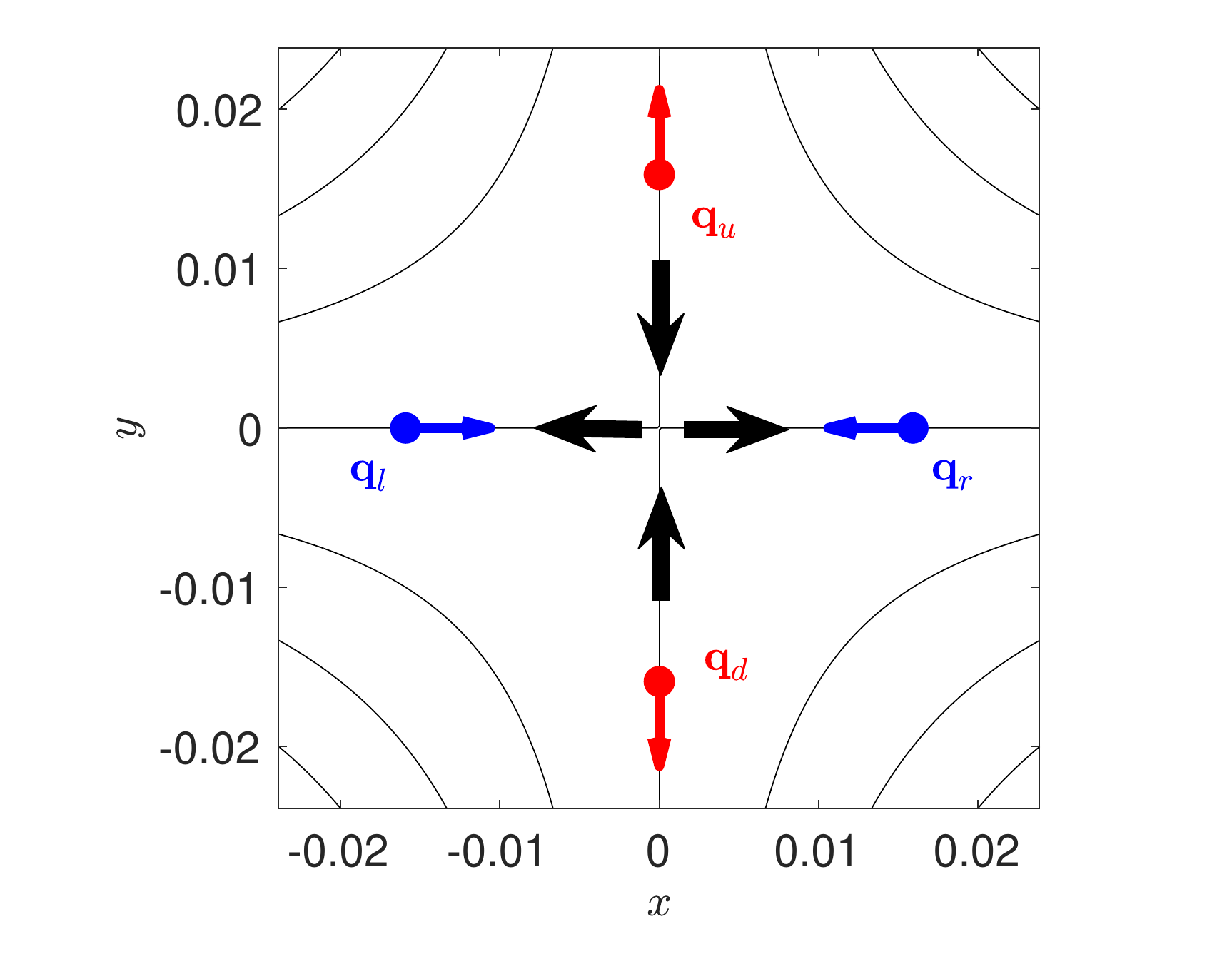}
\caption{Swimming fixed points around the hyperbolic passive fixed point for the linear fluid flow given in Eqs.~\eqref{eq:model_hyperbolic}, with $v_0=0.1$. The black arrows show the direction of the fluid flow, while the colored arrows indicate the orientation of the swimmer (i.e.\ direction of $\hat{\bn}$) at the swimming fixed points. The streamlines of the fluid flow are plotted as solid curves. }\label{fig:hyperbolic_equilibria}
\end{figure}
The analogues to the hyperbolic fixed points of the passive tracer equations are the fixed points of Eqs.~\eqref{eq:model_vortex}.
We refer to the equilibria of Eqs.~\eqref{eq:model_vortex} as swimming fixed points, to distinguish them from the fixed points of passive particles in the fluid flow.
When $v_0 \leq 1$, Eqs.~\eqref{eq:model_vortex} possess at least four equilibria in the vicinity of each hyperbolic fixed point of the fluid flow.
To illustrate the origin of these swimming fixed points, we consider a linear approximation to $\bu(\br)$ in the vicinity of the hyperbolic fixed point at the origin.
This yields
\begin{subequations}\label{eq:model_hyperbolic}
\begin{align} \label{eq:xdot_hyp}
\dot{x} & = 2\pi x + v_0 \cos \theta, \\ \label{eq:ydot_hyp}
\dot{y} & = -2 \pi y+ v_0 \sin \theta, \\ \label{eq:thdot_hyp}
\dot{\theta} & = -2 \pi  \alpha  \sin (2\theta),
\end{align}
\end{subequations}
to first order in $x$ and $y$.
We search for fixed points of Eqs.~\eqref{eq:model_hyperbolic}, i.e.\ phase-space points $\bq$ where $\dot{\bq}=0$.
Equation \eqref{eq:thdot_hyp} implies that $\theta=0, \pi/2, \pi,$ or $3\pi/2$, and for each of these values of $\theta$ it is straightforward to solve Eqs.~\eqref{eq:xdot_hyp} and \eqref{eq:ydot_hyp} for the corresponding $\br$.
This leads to four swimming fixed points, $\bq_l,\bq_r, \bq_u,$ and $\bq_d$, located at a distance $v_0/2\pi$ from the origin, as illustrated in Fig.~\ref{fig:hyperbolic_equilibria}.
Thus, as $v_0 \rightarrow 0$, each of the swimming fixed points approaches the passive hyperbolic fixed point.

The corresponding swimming fixed points in the nonlinear case, i.e.\ for the full flow field given by Eqs.~\eqref{eq:model_vortex}, are straightforward to calculate and they are given by 
\begin{subequations}\label{eq:primary_equilibria}
\begin{align}
\bq_l & = (-\rho,0,0), \\
\bq_r & = (\rho,0,\pi), \\
\bq_u & = \left(0,\rho,\frac{\pi}{2}\right), \\
\bq_d & = \left(0,-\rho,-\frac{\pi}{2}\right),
\end{align}
\end{subequations}
with $\rho = \left(\sin^{-1} v_0\right) / 2\pi$.
We shall refer to these as the primary equilibria of Eqs.~\eqref{eq:model_vortex}.
They are distinguished by their convergence to the hyperbolic passive fixed point in the limit $v_0 \rightarrow 0$, and their direct correspondence to the swimming fixed points of the linearized fluid flow in the vicinity of the hyperbolic fixed point.
Indeed, because the form of Eqs.~\eqref{eq:model_hyperbolic} applies in the vicinity of a hyperbolic fixed point of any  fluid flow, we expect primary swimming equilibria to be generic for swimmers  with a small enough $v_0$ in arbitrary steady incompressible fluid flows with unstable stagnation points.
Notably, the locations of the primary equilibria are independent of $\alpha$; however, the linear stability of the equilibria depends on $\alpha$.
We note that the fixed points come in two pairs, $(\bq_l,\bq_r)$ and $(\bq_u,\bq_d)$.
Within each pair, the fixed points are related to each other by a rotation about the origin by $\pi$, while the two pairs are related to each other by a $t$-symmetry reflection about the lines $y=x$ or $y=-x$.

The linear stability is also straightforward to compute.
The stability matrix $A(\bq) \equiv \partial \dot{\bq} / \partial \bq$ evaluated at the upper equilibrium is 
\begin{equation*}
A(\bq_u) = \begin{pmatrix}
2\pi \sqrt{1-v_0^2} & 0 & -v_0 \\
0 & - 2\pi \sqrt{1-v_0^2} & 0 \\
4\pi^2 v_0 & 0 & 4\pi\alpha \sqrt{1-v_0^2}
\end{pmatrix},
\end{equation*}%NOTES DATE 5/16/2019
where we have used $\cos (2\pi\rho) = \sqrt{1-v_0^2}$.
Diagonalizing this matrix, we obtain the eigenvalues
\begin{subequations}\label{eq:rl_eig}
\begin{align}
\lambda^1_{ud} & = -2\pi \sqrt{1-v_0^2}, \\ \label{eq:rl_eig2}
\lambda^\pm_{ud} & = \pi \left[ (1+2\alpha)\sqrt{1-v_0^2} \pm \sqrt{(1-v_0^2)(1-2\alpha)^2 - 4 v_0^2} \right].
\end{align}
\end{subequations}
We have used the $ud$ subscript in the above equations because $\bq_d$ is related to $\bq_u$ by the rotational symmetry of the equations of motion, so their eigenvalues are identical.
Furthermore, because the left-right pair of eigenvalues is related to the up-down pair by a reversing symmetry, the eigenvalues of $\bq_l$ and $\bq_r$ are given by $\lambda^1_{lr} = - \lambda^1_{ud}$ and $\lambda^\pm_{lr} = - \lambda^\pm_{ud}$.
Since $\lambda^1 _{ud} < 0$ for all $\alpha$ and all $v_0 < 1$, the up-down swimming fixed points always have at least one contracting eigenvector, while the left-right fixed points always have at least one expanding eigenvector.
At $v_0 = 1$, $\lambda_{ud}^1 = \lambda_{lr}^1 = 0$.

Besides the primary swimming fixed points, there is another family of swimming fixed points that exists for certain parameter values and has no analogue in the passive tracer case.
To see this, we note that the equilibrium condition $\dot{\br} = 0$ leads to $\bu = -v_0 \hat{\bn}$.
This in turn implies that $|\bu|^2 = v_0^2$, and a second equation may be obtained by multiplying the $x$- and $y$-components of the previous equation together.
Doing so, $\bu = -v_0 \hat{\bn}$ can be expressed as
\begin{subequations}
\begin{align}\label{eq:u2}
 v_0^2 = & \sin^2(2\pi x) \cos^2(2\pi y) + \cos^2(2\pi x) \sin^2(2\pi y), \\ \label{eq:sin2th}
\sin 2\theta = & -\frac{2}{v_0^2} \sin (2 \pi x) \sin (2\pi y) \cos (2 \pi x) \cos (2 \pi y).
\end{align}
\end{subequations}
Now, we can substitute Eq.~\eqref{eq:sin2th}  into Eq.~\eqref{eq:thdot} to eliminate the $\theta$-dependence from $\dot{\theta} = 0$, leading to 
\begin{equation*}
\sin(2\pi x)\sin(2 \pi y) \left[1 + \frac{2 \alpha}{v_0^2} \cos^2 (2\pi x) \cos^2 (2\pi y) \right] = 0.
\end{equation*}
All swimming fixed points  of Eqs.~\eqref{eq:model_vortex} must satisfy the condition above.
For the primary swimming fixed points, this condition is satisfied due to the fixed points lying on the $x$- or $y$-axes (or half-integer shifts of these axes).
Alternatively, it may be satisfied by having 
\begin{equation}\label{eq:alternativeCond}
2 \cos^2 (2\pi x) \cos^2 (2\pi y) = -\frac{v_0^2}{\alpha}.
\end{equation}
Because the left-hand side is non-negative, this can only occur if $\alpha < 0$, i.e.\ for perpendicular swimmers. 
Whenever points $\br^*$ can be found that simultaneously satisfy Eqs.~\eqref{eq:alternativeCond} and \eqref{eq:u2}, then there are swimming fixed points $\bq^* = (\br^*,\theta^*)$, with the angle $\theta^*$ determined by Eq.~\eqref{eq:sin2th} evaluated at $\br^*$ [modulo $\pi$; this ambiguity can be removed by choosing the $\theta^*$ such that $\hat{\bn}$ is in the opposite direction of $\bu(\br^*)$].
This pair of equations may be solved analytically, leading to
\begin{subequations}\label{eq:secondary_equilibria}
\begin{align}
 \br^*_\pm & = (r_\pm, r_\mp),\,\,\, {\rm with } \\
r_\pm & = \frac{1}{2\pi} \cos^{-1}\left[\left(1 - \alpha \mp \sqrt{(1-\alpha)^2 + \frac{2 \alpha}{v_0^2}}\right)^{-\frac{1}{2}}\right].
\end{align}
\end{subequations}
These expressions correspond to the pair of equilibria in the lower-left corner of the vortex cell  with $\br^* \in [0, \frac{1}{4}]\times [0, \frac{1}{4}]$, and the ``$+$'' (``$-$") sign gives the equilibrium above (below) the $y=x$ $t$-symmetry axis. 
Note that these equilibria $\bq^*$, which we shall call secondary equilibria, depend explicitly on $v_0$ and $\alpha$, as seen in Eqs.~\eqref{eq:secondary_equilibria}.
In contrast, the positions of the primary equilibria only depend on $v_0$ [see Eqs.~\eqref{eq:primary_equilibria}].

\begin{figure}
\includegraphics[width=0.9\textwidth]{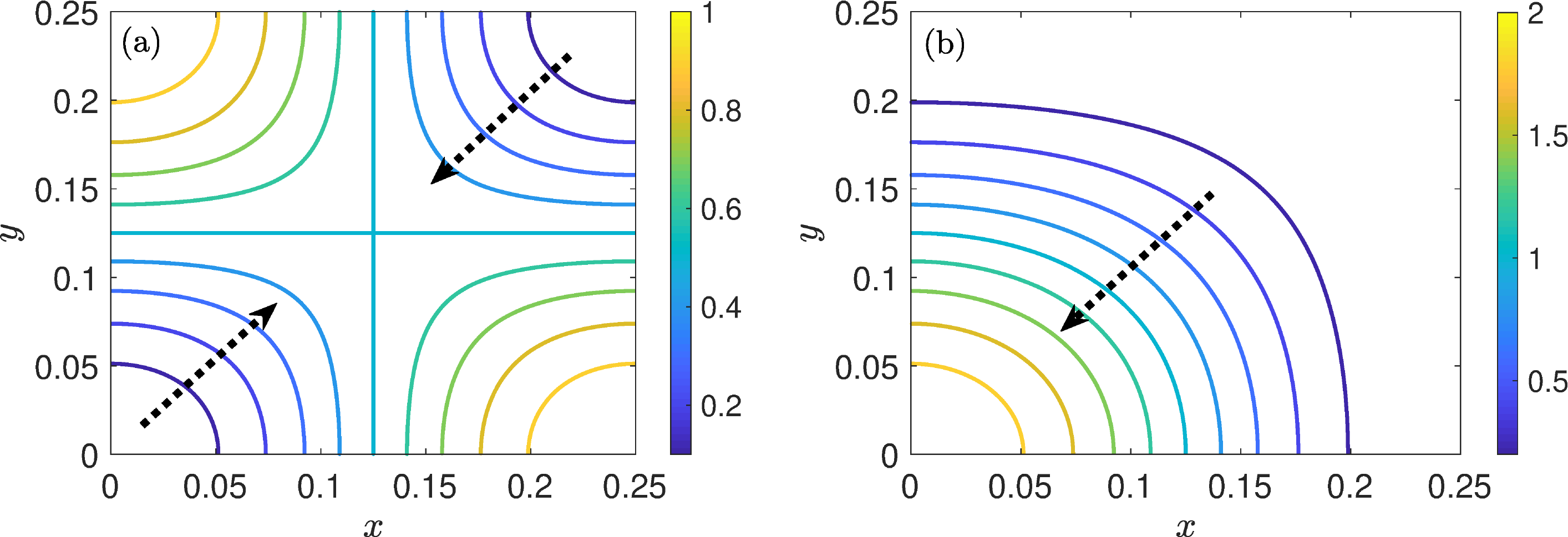}
\caption{Level sets of the functions defining the secondary swimming fixed points. The plots are restricted to the bottom-left quadrant of a single vortex cell. (a) $|\bu|^2$, i.e.\ the right-hand side of Eq.~\eqref{eq:u2}. (b) $2 \cos^2 (2\pi x) \cos^2 (2\pi y)$, left-hand side of Eq.~\eqref{eq:alternativeCond}. Arrows indicate the direction of increasing $v_0$.}\label{fig:equilibria_contours}
\end{figure}
\begin{figure}
\centering
\includegraphics[width=\textwidth]{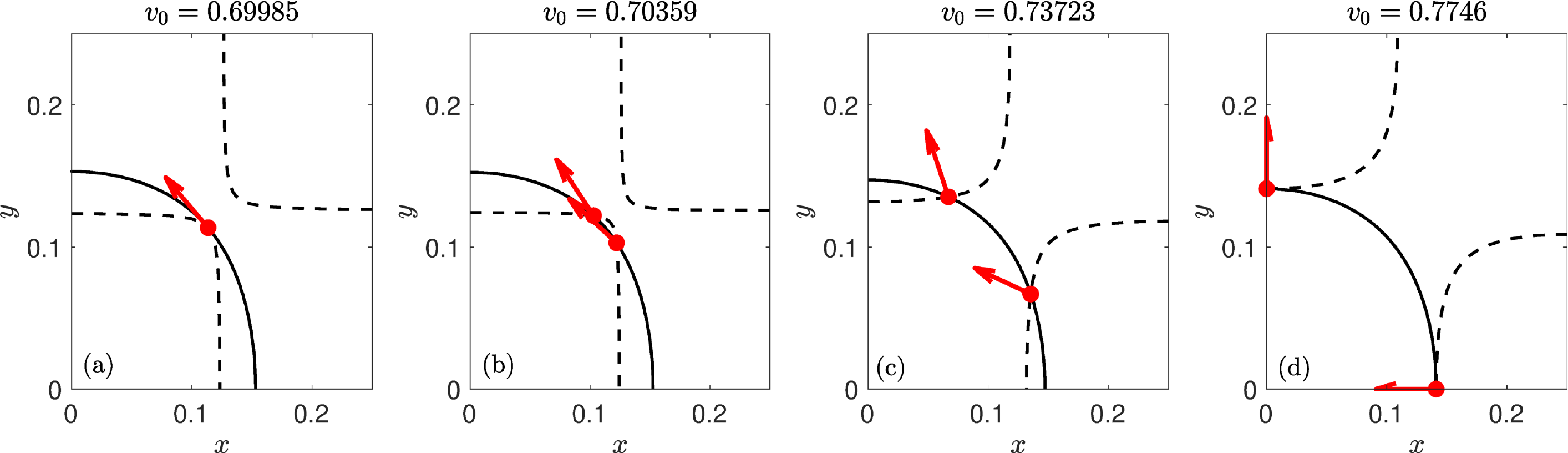}
\caption{Existence of secondary swimming fixed points at $\alpha = -0.75$ with increasing $v_0$, from birth (a) to bifurcation (b) to increasing separation (c) to collision with the primary swimming fixed points (d). The solid curve is defined by Eq.~\eqref{eq:alternativeCond}, while the dashed curves are defined by Eq.~\eqref{eq:u2}. The red dots are the secondary swimming fixed points, and the arrows are the corresponding swimmer orientation. }\label{fig:secondary_equilibria}
\end{figure}

The emergence of the secondary equilibria can be understood geometrically, by considering the curves in the $xy$ plane defined by Eqs.~\eqref{eq:u2} and \eqref{eq:alternativeCond}.
For a given $v_0$ and $\alpha$, these curves are level sets of the functions on the right- and left-hand sides of Eqs.~\eqref{eq:u2} and \eqref{eq:alternativeCond}, respectively, as shown in Fig.~\ref{fig:equilibria_contours}.
The intersection between these curves, if it occurs, is precisely at the points $\br^*$.
Focusing near the origin, Fig.~\ref{fig:equilibria_contours}a illustrates that as $v_0$ increases, the corresponding level curve of $|\bu|^2$ expands outward from the origin.
Meanwhile, for a fixed $\alpha$, the corresponding level curve of the left-hand side of Eq.~\eqref{eq:alternativeCond} shrinks inward to the origin as $v_0$ increases.
Eventually, the two curves intersect at a point along the symmetry axis $y=x$, as illustrated in Fig.~\ref{fig:secondary_equilibria}a for $\alpha = -0.75$, and a $t$-symmetric secondary swimming fixed point is born.
As $v_0$ is increased further, this fixed point bifurcates into two secondary swimming fixed points, connected by the $t$-symmetry about the $y=x$ axis (Fig.~\ref{fig:secondary_equilibria}b).
The fixed points move away from each other (Fig.~\ref{fig:secondary_equilibria}c) until, at a critical $v_0$, the secondary swimming fixed points collide with the primary swimming fixed points on the $x$- and $y$-axes, and subsequently disappear (Fig.~\ref{fig:secondary_equilibria}d).
Note that the change in topology of the level curves of $|\bu|^2$ going from Fig.~\ref{fig:secondary_equilibria}b to Fig.~\ref{fig:secondary_equilibria}c is not critical to the formation or persistence of the secondary swimming fixed points.

Next, we illustrate how the collision of the secondary equilibria with the primary equilibria is related to the stability of the primary equilibria.
First, we explicitly compute the bifurcation curves for the secondary equilibria.
As shown above, the secondary fixed points are born when Eqs.~\eqref{eq:alternativeCond} and \eqref{eq:u2} are simultaneously satisfied at a point such that $y^*=x^*$.
Imposing this condition, using Eq.~\eqref{eq:alternativeCond} to eliminate $x^*$ from Eq.~\eqref{eq:u2}, and solving for $v_0$ in terms of $\alpha$, we obtain
\begin{equation}\label{eq:vbirth}
v_{\rm birth}(\alpha) = \sqrt{\frac{2 \alpha}{2 \alpha - 1 - \alpha^2}}.
\end{equation}
On the other hand, the secondary swimming fixed points are destroyed when they collide with one of the primary swimming fixed points, for example $\bq_u = \bq_+^*$.
Here, this gives $x^*_+=0$, and $y^*_+ = \rho$.
Eliminating $y^*_+$ from the previous equation using Eq.~\eqref{eq:alternativeCond} and solving for $v_0$ in terms of $\alpha$ yields
\begin{equation}\label{eq:vdeath}
v_{\rm death}(\alpha) = \sqrt{\frac{2 \alpha}{2 \alpha - 1}}.
\end{equation}
Hence, the secondary swimming fixed points exist for parameters such that $v_{\rm birth}(\alpha) \leq v_0 < v_{\rm death}(\alpha)$.

\begin{figure}
\centering
\includegraphics[width=0.8\textwidth]{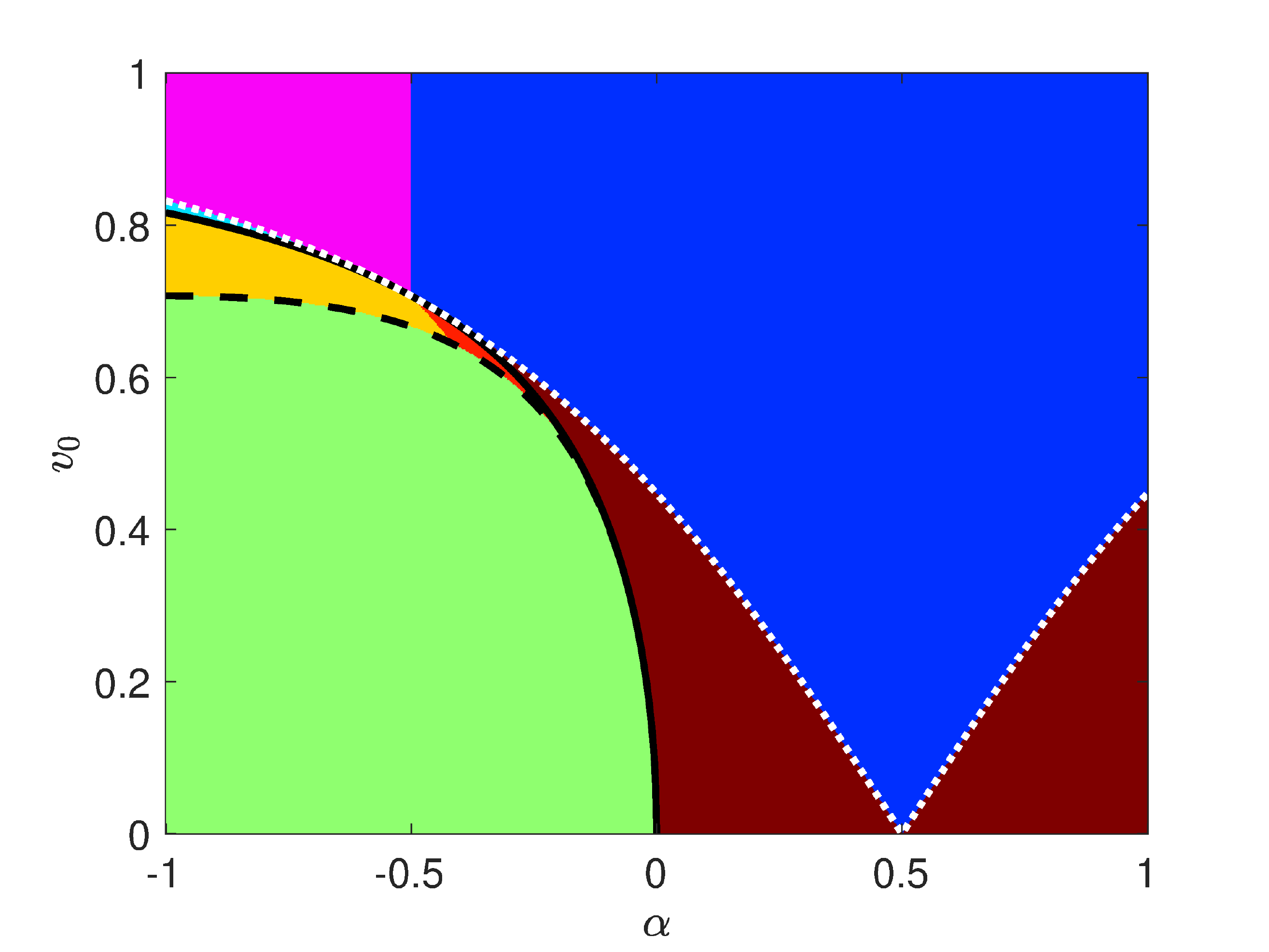}
\caption{(color online) Stability diagram of the primary swimming fixed points $\bq_{ud}$ and secondary swimming fixed point $\bq_+^*$. The dashed black curve demarcates $v_{\rm birth}(\alpha)$, the birth of the secondary swimming fixed points with increasing $v_0$. The solid black curve demarcates $v_{\rm death}(\alpha)$, the collision of the secondary swimming fixed points with the primary swimming fixed points.  The dotted white curve demarcates $v_c(\alpha)$, the transition from real to complex eigenvalues for the primary swimming fixed points. Each color corresponds to a different combination of stability types of the $\bq_{ud}$ and $\bq_+^*$ fixed points, when the latter exists. Magenta: $\bq_{ud}$ SSS complex. Blue: $\bq_{ud}$  SUU complex. Light blue: $\bq_{ud}$  SSS real. Yellow: $\bq_{ud}$  SSU real, $\bq_+^*$ SSS. Red: $\bq_{ud}$ SSU real, $\bq_+^*$ SUU.  Green: $\bq_{ud}$ SSU real. Dark red: $\bq_{ud}$ SUU real. }\label{fig:equilbria_stabilty}
\end{figure}
In Figure \ref{fig:equilbria_stabilty}, we have plotted $v_{\rm birth}(\alpha)$ and $v_{\rm death}(\alpha)$, along with the stability types of the $\bq_{ud}$ and $\bq_{+}^*$ swimming fixed points.
For $\bq_{ud}$, using Eqs.~\eqref{eq:rl_eig} we find that there are five distinct possibilities, as summarized in Table \ref{tab:equilibria_stability}.
Because the sign of the eigenvalue $\lambda^1_{ud}$ is fixed, the stability of the primary equilibria depends on $\operatorname{Re}[\lambda_\pm]$.
When an eigenvalue $\lambda$ has $\operatorname{Re}[\lambda] > 0$, it contributes an unstable direction (U), while if  $\operatorname{Re}[\lambda] < 0$  then it contributes a stable direction (S).
Also, because $\lambda_{lr} = -\lambda_{ud}$, an unstable (stable) direction of an up-down fixed point becomes a stable (unstable) direction of the corresponding left-right fixed point.
The eigenvalues may all be purely real, or they may contain a complex-conjugate pair when the argument of the square-root in Eq.~\eqref{eq:rl_eig2} goes negative.
This occurs when $v_0 > v_c(\alpha)$, with
\begin{equation*}
v_c(\alpha) = \begin{cases}
&- (2\alpha - 1)\left[(2\alpha-1)^2 + 4)\right]^{-1/2} \,\,\,{\rm for}\,\, -1 \leq \alpha \leq \frac{1}{2}, \\
 & (2\alpha - 1)\left[(2\alpha-1)^2 + 4)\right]^{-1/2} \,\,\,{\rm for}\,\,  \frac{1}{2} < \alpha \leq 1.
 \end{cases}
\end{equation*}
On the other hand, we compute the stability of $\bq_{+}^*$ numerically.
We find that the stability type is either SSS real, SSS complex, SUU real, or SUU complex; however, in Fig.~\ref{fig:equilbria_stabilty}, we do not distinguish between the real and complex variants for clarity.
We see in Fig.~\ref{fig:equilbria_stabilty} that the destruction of the secondary equilibria as $v_0$ increases at a fixed $\alpha$ coincides with a change of stability of the $\bq_{ud}$ equilibria from SSU to either SSS real (small light blue sliver) for $\alpha < -\frac{1}{2}$ or SUU real (dark red) for  $\alpha > -\frac{1}{2}$.
From Table \ref{tab:equilibria_stability}, we see that both of these changes of stability occur when $\lambda_+=0$, and by imposing this condition on Eq.~\eqref{eq:rl_eig2}, we indeed recover Eq.~\eqref{eq:vdeath} for $v_{\rm death}(\alpha)$.
At $\alpha = -\frac{1}{2}$, we have $v_{\rm death} = v_c$, and for $v_0 > v_c$ we observe a transition from SSS complex (magenta) for $\alpha < -\frac{1}{2}$ to SUU complex (blue) for $\alpha > -\frac{1}{2}$ .
We remark that the stability types we see on either side of the collision of the secondary and primary swimming fixed points at $\alpha = -1$ are consistent with the constraints imposed by the topological index properties of fixed points of Eqs.~\eqref{eq:model_general} with $\alpha=-1$,\cite{Mitchell2012} and for the entire range of $\alpha$ they are consistent with the topological index properties of fixed points of general $n$-dimensional dynamical systems. \cite{Hsu1980_2}
\begin{table}
\centering
\begin{tabular}{c | c | c}
$\bq_{ud}$ eigenvalue properties &   $\bq_{ud}$ stability & $\bq_{lr}$ stability \\ \hline
$\operatorname{Im}[\lambda_\pm] \neq 0$, $\operatorname{Re}[\lambda_+] < 0$, $\operatorname{Re}[\lambda_-] < 0$ & SSS complex & UUU complex \\
$\operatorname{Im}[\lambda_\pm] \neq 0$, $\operatorname{Re}[\lambda_+] > 0$, $\operatorname{Re}[\lambda_-] > 0$ & SUU complex & SSU complex \\
$\operatorname{Im}[\lambda_\pm] = 0$, $\operatorname{Re}[\lambda_+] < 0$, $\operatorname{Re}[\lambda_-] <  0$ &  SSS real  & UUU real \\
$\operatorname{Im}[\lambda_\pm] = 0$, $\operatorname{Re}[\lambda_+] > 0$, $\operatorname{Re}[\lambda_-] < 0$ & SSU real  & SUU real\\
$\operatorname{Im}[\lambda_\pm] = 0$, $\operatorname{Re}[\lambda_+] > 0$, $\operatorname{Re}[\lambda_-] > 0$ & SUU real & SSU real
\end{tabular}
\caption{Linear stability properties of the primary swimming fixed points.}\label{tab:equilibria_stability}
\end{table}

\begin{figure}
\centering
\includegraphics[width=0.7\textwidth]{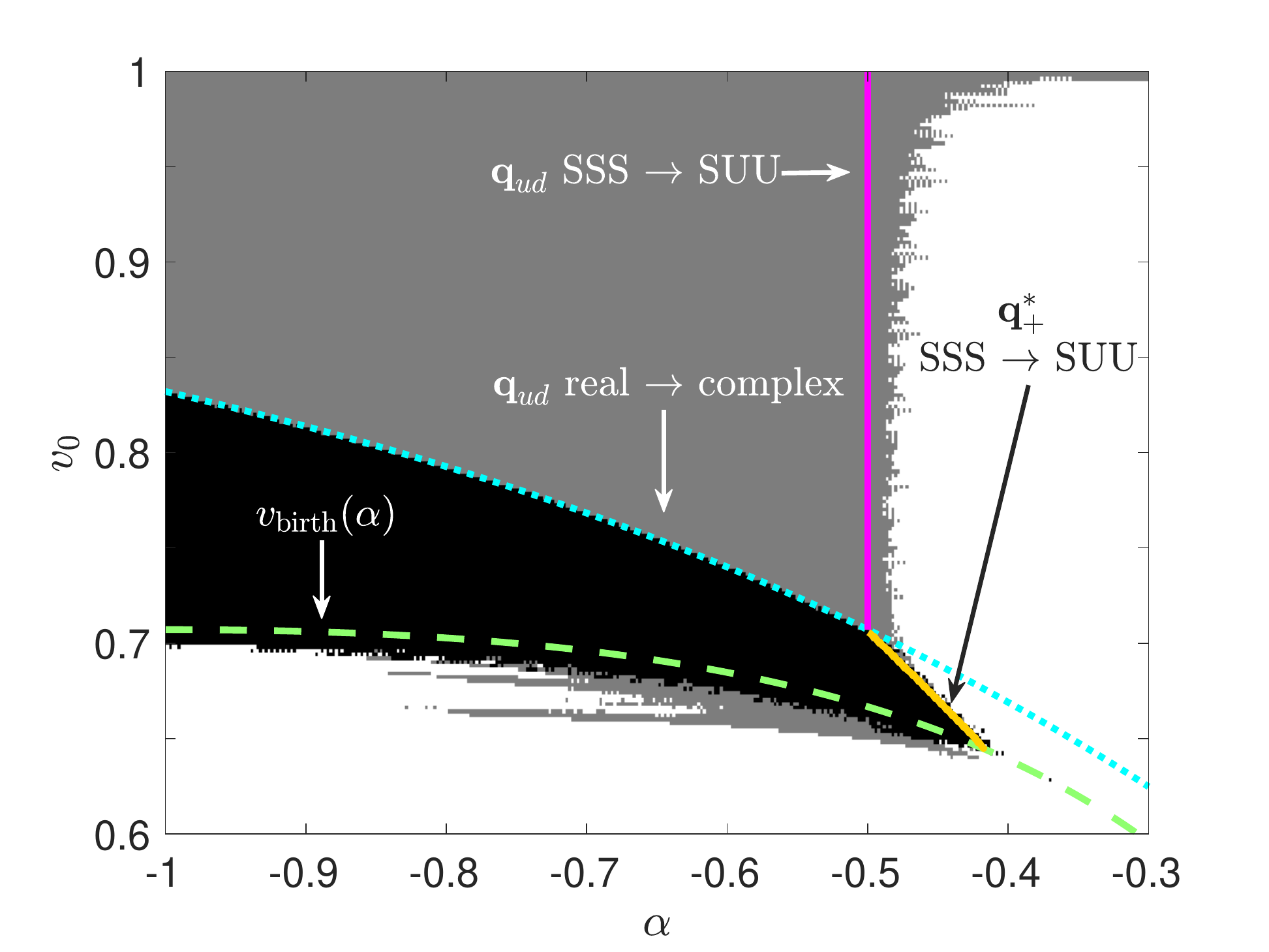}
\caption{High-$v_0$ trapping and localization of perpendicular swimmers as a function of $v_0$ and $\alpha$. White regions indicate no localization or trapping, black regions indicate nonzero trapping, and grey regions indicate nonzero localization without trapping. Selected swimming fixed point bifurcations are plotted as the labeled, colored curves.}\label{fig:eq_and_stab}
\end{figure}

Swimming fixed points with SSS stability type are asymptotically stable, and hence we expect nearby swimmers to become trapped (or localized) in their vicinity.
Indeed, the existence of asymptotically stable swimming fixed points accounts for the high-$v_0$ trapping and localization seen for perpendicular swimmers in Figs.~\ref{fig:monte_carlo} and \ref{fig:monte_carlo_loc}, respectively.
This is illustrated in Fig.~\ref{fig:eq_and_stab}, where we have superimposed the swimming fixed point bifurcations in which asymptotic stability is lost over a magnification of the high-$v_0$ localization region, showing where either trapping or localization only are nonzero.
Evidently, for high $v_0$, localization and trapping are mostly confined to the regions of parameter space where asymptotically stable swimming fixed points exist.
For most values of $v_0$, the probability of trapping and localization quickly drops to zero outside of these regions.
We suspect that the drop-off is not completely abrupt due to finite-time trapping of swimmers, for times exceeding the integration time of $t_f=40$ of our simulations.
One cause of this may be swimming fixed points with eigenvalues with near-zero real parts, which is the case near the $\bq_{ud}$ SSS $\rightarrow$ SUU transition, as well as near $v_0 = 1$ for all values of $\alpha$ [see Eqs.~\eqref{eq:rl_eig}].
Alternatively, for parameters near the creation of an SSS equilibrium, e.g.\ just below $v_{\rm birth}(\alpha)$ in Fig.~\ref{fig:eq_and_stab}, there may be a phase-space bottleneck through which trajectories escape very slowly. \cite{Strogatz}

We also note that trapping at high $v_0$ only occurs when there are either SSS secondary swimming fixed points or SSS real primary swimming fixed points.
In particular, the sharp transition from trapping to localization without trapping in Fig.~\ref{fig:eq_and_stab} coincides with the $\bq_{ud}$ real $\rightarrow$ complex transition.
This may be understood from the following arguments.
The secondary swimming fixed points are strictly inside the vortex cell (see Fig.~\ref{fig:secondary_equilibria}), so swimmers initially close enough to them will remain inside the vortex cell.
When localization is due to the primary swimming fixed points, which are on the boundary of the vortex cell, a swimmer can only be trapped (i.e.\ confined strictly within the cell) for all time if all the eigenvalues are real.
Once the eigenvalues become complex, swimmers must spiral into the fixed point, and the spiraling necessarily causes crossing the vortex cell boundary.
Hence, these swimmers are not considered trapped, but they are still localized.
 
To summarize, we have identified all swimming fixed points of Eqs.~\eqref{eq:model_vortex} for all parameters $v_0$ and $\alpha$, and we have shown that they become asymptotically stable through certain bifurcations, which leads to the high-$v_0$ trapping of perpendicular swimmers observed in our numerical simulations.
In the next section, we identify the invariant solutions of Eqs.~\eqref{eq:model_vortex} that account for swimmer trapping at low-to-intermediate values of $v_0$, namely, periodic orbits and invariant tori.
The swimming fixed points discussed here will also play a role: they are involved in global bifurcations of the periodic orbits.
%As noted in Table \ref{tab:equilibria_stability}, the UUU stability of the primary swimming fixed points $\bq_{lr}$  for $v_0 > v_{\rm death}(\alpha)$ and $\alpha <- \frac{1}{2}$ implies the SSS stability of $\bq_{ud}$ due to $t$-symmetry.
%Hence, for this range of parameters we conclude that there are linearly stable, attracting swimming fixed points on the boundary of the vortex cells for sufficiently elongated perpendicular swimmers.
%This is surprising, since this only occurs for quite high swimming speeds $v_0 \gtrsim 0.7$.
%In principle, this property may lead to the trapping of swimmers near (and asymptotically approaching) these fixed points, though this mechanism must be distinct from that discussed in Sec.~\ref{sec:numerics}, where our attention is restricted to swimmer trapping strictly inside the interior of a vortex cell.
%We have indeed observed such swimmer trajectories in isolated numerical experiments in the $v_0 > v_{\rm death}(\alpha)$ and $\alpha <- \frac{1}{2}$ parameter range, but a systematic investigation of this phenomenon is outside the scope of this paper.

\section{Periodic orbits}\label{sec:po}
A natural question is whether the previous analysis can be extended to the region near the vortex center, in order to identify swimming fixed point analogues of the passive elliptic fixed points.
However, a linearization of the fluid flow in the vicinity of a passive elliptic fixed point applied to Eqs.~\eqref{eq:model_vortex} reveals that there are no swimming fixed points in that region, as we show in the next section and is already implied by the results of Sec.~\ref{sec:eq}.
On the other hand, a simplified nonlinear analysis of the flow in this region for small $v_0$ suggests the existence of a  stable periodic orbit surrounded by a continuous family of quasi-periodic orbits, i.e. invariant tori, which trap swimmers inside a vortex cell indefinitely.
In this section, we develop this simplified analysis and then show numerically that this stable periodic orbit and family of tori also exists for the full flow field of Eqs.~\eqref{eq:model_vortex}.
It is in fact the stable periodic orbit seen at the center of the family of tori in Fig.~\ref{fig:poincare}.
We then investigate the bifurcations of this periodic orbit with increasing $v_0$ and propose a criterion for the breakdown of trapping in the low-to-intermediate $v_0$ range.

\subsection{Near the vortex center: small $v_0$ limit}
To understand the swimmer trajectories near the center of a vortex, we expand the flow about the center.
First, we shift our coordinates to the center of a vortex cell: $\tilde{\br} = \br - (1/4,1/4)$.
In these coordinates, the stream function becomes $\psi(\tilde{\br}) = \cos(2\pi \tilde{x}) \cos(2 \pi \tilde{y})/2\pi$.
Then we move to polar coordinates $(\tilde{x},\tilde{y}) = (r\cos\phi, r\sin \phi)$ and we expand $\psi$ assuming $r$ is small, leading to
\begin{equation}\label{eq:streamPolar}
\psi(r,\phi) = -\pi r^2 + \frac{\pi^3}{2}r^4 - \frac{\pi^3}{6} r^4 \cos (4\phi) + O(r^6),
\end{equation}
where we have neglected a constant.
In polar coordinates, the fluid velocity is related to the stream function as
\begin{equation}
\bu(r,\phi) = \frac{\psi_{,\phi}}{r} \hat{\br} + r \Omega \,\hat{\pmb{\phi}},\,\,\,{\rm with}\,\, \Omega = -\frac{\psi_{,r}}{r}.
\end{equation}
Here, $\Omega$ is the instantaneous passive tracer rotation frequency. 
We note that up to $O(r^6)$, Eq.~\eqref{eq:streamPolar} can be broken up into a circularly-symmetric part---the first two, $\phi$-independent terms---and a part explicitly dependent on $\phi$.

For the case of circularly-symmetric flows, i.e.\ when $\psi_{,\phi} = 0$, the passive tracer equations of motion would be
\begin{align*}
\dot{r} & = 0, \\
\dot{\phi} & = \Omega(r),
\end{align*}
meaning fluid particles move on circular orbits at a fixed frequency $\Omega(r)$.
For swimmers, the corresponding equations of motion are \cite{Torney2007}
\begin{subequations}\label{eq:model_circ_vortex}
\begin{align}
\dot{r} & = v_0 \cos(\theta - \phi), \\
\dot{\phi} & = \Omega(r)+ \frac{v_0 \sin (\theta - \phi)}{r}, \\
\dot{\theta} & = \Omega(r) + \frac{1}{2}r \Omega_{,r} \left(1 + \alpha \cos[2(\theta-\phi)]\right).
\end{align}
\end{subequations}
Assuming $r$ is very small and truncating  $O(r^4)$ terms and higher from Eq.~\eqref{eq:streamPolar}, we obtain a circularly-symmetric flow with $\Omega(r) = 2\pi$.
In fact, this corresponds to a linearization of the fluid flow about the elliptic equilibrium point in Cartesian coordinates.
In polar coordinates, the signature of the linearization is the fact that $\Omega$ is constant, independent of $r$.
Adding in the swimmer motility, the swimmer equations of motion in this flow are
\begin{subequations}\label{eq:circ_symm}
\begin{align}
\dot{r} & = v_0 \cos(\theta - \phi), \\
\dot{\phi} & = 2\pi + \frac{v_0 \sin (\theta - \phi)}{r}, \\
\dot{\theta} & = 2\pi.
\end{align} 
\end{subequations}
Clearly, there can be no swimming fixed points because $\dot{\theta} \neq 0$.
An analysis in Cartesian coordinates (where the equations are linear) shows that all trajectories are eventually dominated by the particular solution $(r,\phi,\theta) = (v_0t, 2 \pi t + \theta_0, 2\pi  t +\theta_0)$, for a constant $\theta_0$.
In other words, the linear theory predicts that all swimmers eventually spiral out from the center of the vortex by swimming radially outward.
Furthermore, this result is independent of $\alpha$ because this parameter is absent from Eqs.~\eqref{eq:circ_symm}.

\begin{figure}
\centering
\includegraphics[width=\textwidth]{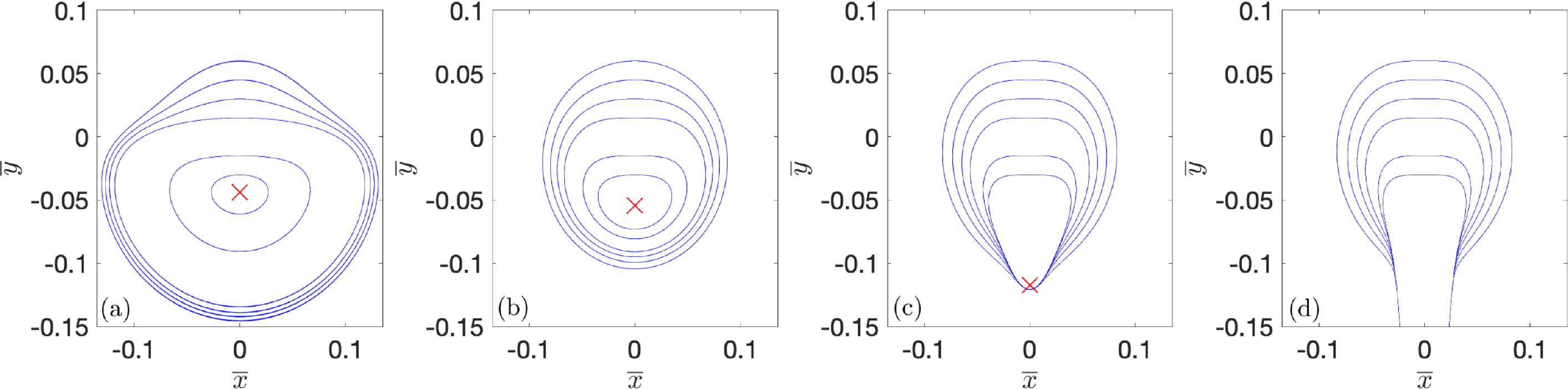}
\caption{Phase portrait for a swimmer in a nonlinear, circularly symmetric vortex, using symmetry-reduced coordinates $(\overline{x},\overline{y}) = (r\cos\beta, r\sin\beta)$. The parameters are $v_0 = 0.01$ and $\alpha = -0.9$ (a), $\alpha = 0$ (b),  $\alpha = 0.9$ (c), and $\alpha = 1$ (d). The red ex indicates the stable equilibrium Eq.~\eqref{eq:equilibrium_approx}.}\label{fig:symm_phase_portrait}
\end{figure}
As shown in Ref.~\onlinecite{Torney2007}, however, taking into account the nonlinearity through an $r$-dependent $\Omega$ leads to the formation of stable swimmer orbits near the vortex center, on which swimmers are trapped indefinitely.
Following Ref.~\onlinecite{Torney2007}, we reduce the rotational symmetry of Eqs.~\eqref{eq:model_circ_vortex} by going into a rotating frame, where the  dynamical variables are $r$ and $\beta \equiv \theta - \phi$.
Hence, the system is reduced to two equations,
\begin{subequations}\label{eq:model_circ_vortex_rot}
\begin{align}
\dot{r} & = v_0 \cos\beta, \\ \label{eq:betadot}
\dot{\beta} & =  - \frac{v_0 \sin \beta}{r} + \frac{1}{2}r \Omega_{,r} \left[1 + \alpha \cos(2\beta)\right],
\end{align}
\end{subequations}
which do not depend on $\phi$.
Now, the full flow field up to $O(r^4)$ is not circularly symmetric because of the last term in Eq.~\eqref{eq:streamPolar}.
Nevertheless, this term is always smaller in magnitude than the preceding terms, and its value averaged over $\phi$ is zero.
Thus, it is plausible to neglect the $\phi$-dependent term as a first approximation, and consider the resulting flow with
$\Omega(r) = 2\pi - 2\pi^3 r^2$.
Hence, Eqs.~\eqref{eq:model_circ_vortex_rot} become
\begin{subequations}\label{eq:model_approx}
\begin{align}
\dot{r} & = v_0 \cos\beta, \\ 
\dot{\beta} & =  - \frac{v_0 \sin \beta}{r} -2\pi^3 r^2 \left[1 + \alpha \cos(2\beta)\right].
\end{align}
\end{subequations}

The phase portrait for these equations, computed through the numerical integration of Eqs.~\eqref{eq:model_approx}, is shown in Fig.~\ref{fig:symm_phase_portrait}, where we have introduced the coordinates $(\overline{x},\overline{y}) = (r\cos\beta, r\sin\beta)$ for visualization purposes.
For a fixed $v_0$ and all $\alpha < 1$, the structure of phase space is a continuous family of nested periodic orbits surrounding the unique equilibrium of Eqs.~\eqref{eq:model_approx}.
This equilibrium is located at
\begin{equation}\label{eq:equilibrium_approx}
(r^*,\beta^*) = \left(\frac{1}{\pi}\sqrt[3]{\frac{v_0}{2(1-\alpha)}},\frac{3\pi}{2}\right),
\end{equation}
and its eigenvalues are given by
\begin{equation*}
\lambda_\pm =\pm i \sqrt{v_0\left[\frac{v_0}{(r^*)^2} + 4\pi^3 r^*(1 - \alpha)\right]}.
\end{equation*}
Because the eigenvalues are purely imaginary, the equilibrium is linearly stable.
Furthermore, it is also invariant under the $t$-symmetry of Eqs.~\eqref{eq:model_approx} with the involution $(r,\beta) \mapsto (r,\pi-\beta)$.
Therefore, the equilibrium is also nonlinearly stable and surrounded by periodic orbits (at least locally),\cite{Strogatz,Lamb1998} as confirmed by the phase portraits in Figs.~\ref{fig:symm_phase_portrait}a--c.

Hence for $\alpha < 1$, all swimmers in the circularly symmetric, nonlinear vortex flow given by Eqs.~\eqref{eq:model_approx} are trapped on bounded orbits. 
When moving into the full $(r,\phi,\theta)$ phase space, the equilibrium \eqref{eq:equilibrium_approx} becomes a periodic orbit with a period set by the rotation of $\phi$.
Specifically, it is a circular trajectory with radius $r^*$ and the swimmer always oriented in the upstream direction.
This orbit is the analogue of the passive elliptic fixed point.
On the other hand, the periodic orbits of Eqs.~\eqref{eq:model_approx} are in general quasi-periodic orbits in the full phase space, because in general the period in $(r,\beta)$ space is incommensurate with the period of $\phi$.
Hence, the full phase space is foliated by invariant tori surrounding the stable periodic orbit.
In other words, swimmer motion for $\alpha < 1$ in this flow is integrable.
The invariant tori are the analogues of the periodic orbits of the passive tracer system.
If a swimmer is on an invariant torus, its distance from the vortex center oscillates as its swimming direction relative to the local flow oscillates, as seen in Fig.~\ref{fig:symm_phase_portrait}.
At $\alpha = 1$, the equilibrium of Eqs.~\eqref{eq:model_approx} goes to infinity and all the periodic orbits are broken (Fig.~\ref{fig:symm_phase_portrait}d).
Consequently, all swimmers eventually spiral out from the vortex center,\cite{Torney2007} as in the linear case. 

Based on this analysis, we might expect the full system, Eqs.~\eqref{eq:model_vortex}, to possess a $t$-symmetric, stable periodic orbit near the center of each vortex surrounded by a family of invariant tori.
Swimmers on these orbits would thus remain trapped inside individual vortex cells for all time, like passive fluid tracers circulating around the passive elliptic fixed point.
However, the above analysis relied on the assumptions that $r$ is small and the rotational asymmetry of the true vortex flow is negligible.
From Eq.~\eqref{eq:equilibrium_approx}, we see that the first assumption breaks down when either $v_0$ becomes large or $\alpha \rightarrow 1$.
In these limits, we would thus expect the stable periodic orbit to bifurcate and possibly disappear completely.
Indeed, numerical simulations for $0 \leq \alpha \leq 1$ indicate a critical $v_0$ which depends on $\alpha$ above which there is a complete absence of swimmer trapping inside individual vortex cells, suggesting an absence of stable periodic orbits confining a swimmer to a single vortex cell. \cite{Torney2007}
As $\alpha \rightarrow 1$, it appears that this critical $v_0$ approaches zero.
Even if a stable periodic orbit of the full system corresponding to Eq.~\eqref{eq:equilibrium_approx} exists, the above analysis gives no indication of how large $r$ can get such that a quasi-periodic orbit with this $r$ of the full system can be found.
Intuition based on Kolmogorov-Arnold-Moser (KAM) theory for reversible systems \cite{Sevryuk1998} suggests that even in the presence of rotational asymmetry and other perturbations to this model due to increasing $r$, the stable periodic orbit and many of the surrounding invariant tori may persist.
To address these questions, we now turn to the direct numerical computation of the periodic orbit of the full system corresponding to Eq.~\eqref{eq:equilibrium_approx} for the full range of parameters $v_0$ and $\alpha$.

\subsection{Exact periodic orbits for finite $v_0$}
\begin{figure}
\centering
\includegraphics[width=0.6\textwidth]{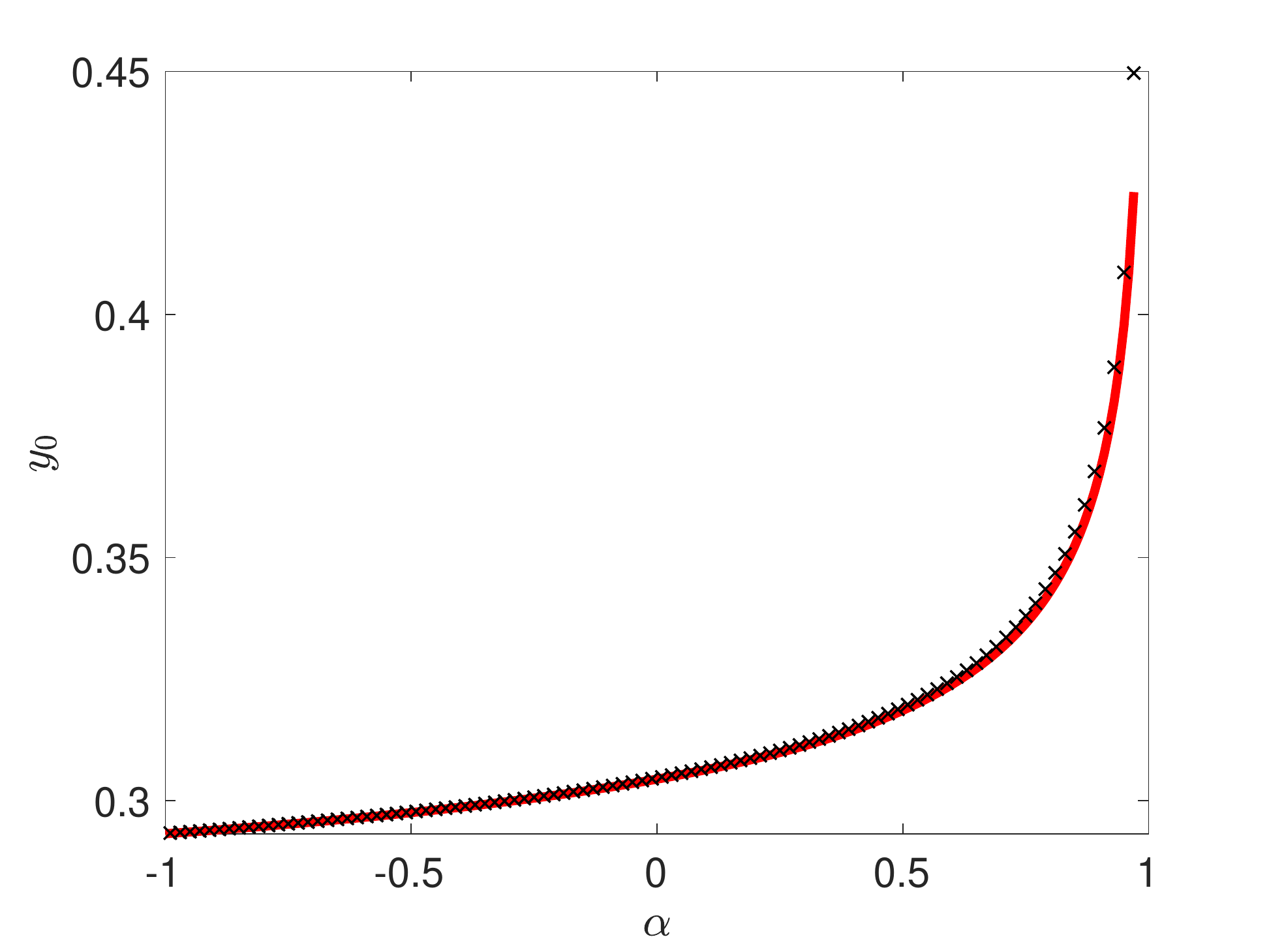}
\caption{Initial condition $y_0$ of the main stable periodic orbit, with $x_0 = 1/4$ and $\theta_0 = 0$, for $v_0 = 0.01$. Black exes: exact initial condition of periodic solution of Eqs.~\eqref{eq:model_vortex}. Solid red curve: Analytical prediction given by Eq.~\eqref{eq:guess}. }\label{fig:main_y0}
\end{figure}
In the limit $v_0 \rightarrow 0$, we expect Eqs.~\eqref{eq:model_vortex} to have a periodic solution with an initial condition determined by Eq.~\eqref{eq:equilibrium_approx}.
Equation \eqref{eq:equilibrium_approx} can be converted into an initial condition $\bq_0$ in $(x,y,\theta)$ coordinates by fixing the initial angle $\phi_0$ about the center of the vortex, and then converting from $(r^*,\beta^*)$ to $(x_0,y_0,\theta_0)$.
Taking $\phi_0=\pi/2$, the initial condition predicted using Eq.~\eqref{eq:equilibrium_approx} is 
\begin{equation}\label{eq:guess}
\bq_0 = \left(\frac{1}{4},\frac{1}{4} + r^*,0\right).
\end{equation}
Because $x_0 = 1/4$ and $\theta_0 = 0$, this initial condition is invariant under the $t$-symmetry of reflection about the vertical axis through the vortex center $\br = (1/4,1/4)$ (see Table \ref{tab:tsymm}).
Therefore, the periodic orbit predicted by Eq.~\eqref{eq:guess} is $t$-symmetric, i.e.\ invariant under the $t$-symmetry. \cite{Roberts1992}

We are indeed able to find an exact, $t$-symmetric periodic solution through the direct numerical integration of Eqs.~\eqref{eq:model_vortex} for a fixed $v_0 = 0.01$ and $-1 \leq \alpha \leq 0.97$.
Here we summarize these results, while details on our procedure for numerically computing the periodic orbits are given in Sec.~\ref{sec:continuation}.
In Fig.~\ref{fig:main_y0}, we compare the predicted initial condition $y_0$ to that of the exact $t$-symmetric periodic orbit of Eqs.~\eqref{eq:model_vortex}.
The agreement is excellent for most values of $\alpha$ except near $1$, where the true $y_0$ begins to significantly exceed the predicted one.
This is also where the true initial condition gets close to the vortex cell boundary at $y = 1/2$, where the rotational symmetry underlying the prediction of Eq.~\eqref{eq:guess} breaks down, so it is not surprising that there is a large disagreement here.
We have checked that the agreement between the two calculations improves for smaller values of $v_0$, confirming the accuracy of Eq.~\eqref{eq:guess} in the limit $v_0 \rightarrow 0$.
Furthermore, the exact periodic orbit possesses additional symmetries: it is  invariant under the $\pi/2$ rotational symmetry about the vortex center and the $y=x$ $t$-symmetry.
Also, it is always of center stability type for this range of parameters.
That is, besides the trivial marginal eigenvalue, the remaining two eigenvalues are complex with unit modulus.
This is consistent with the center stability of the equilibrium  \eqref{eq:equilibrium_approx} in the rotating frame.
Center periodic orbits, which are generic in Hamiltonian systems, also occur generically in reversible dynamical systems. \cite{Roberts1992}
They must be invariant with respect to at least one of the system's $t$-symmetries, which is clearly true in this case.
As in the case of Hamiltonian systems, they also are generically surrounded by families of invariant tori. \cite{Roberts1992}
In particular, the existence of a $t$-symmetric center-type periodic orbit for Eqs.~\eqref{eq:model_vortex} implies the existence of a corresponding family of tori.
Therefore, we have established the existence of a family of quasi-periodic solutions near the main periodic orbit that traps swimmers inside individual vortex cells for all time, in analogy with the periodic orbits surrounding the elliptic fixed point of passive tracers.

As $v_0$ is increased at a fixed $\alpha$, the main periodic orbit eventually bifurcates, either disappearing completely or changing its stability through the creation or destruction of additional periodic orbits.
In the following sections, we investigate these bifurcations through the numerical continuation of the periodic orbits plotted in Fig.~\ref{fig:main_y0} with increasing $v_0$.
We show that these bifurcations allow the accurate prediction of the possibility, or lack thereof, of swimmer trapping up to intermediate swimming speeds.

\subsubsection{Periodic orbit continuation schemes with increasing $v_0$}\label{sec:continuation}
Here, we briefly describe the numerical schemes we use to continue periodic orbits of Eqs.~\eqref{eq:model_vortex} at a fixed $\alpha$ as a parameter is varied.
The automated computation of periodic orbits requires two elements: a root-finding algorithm, and a method for generating good initial guesses for the periodic orbit initial conditions and other parameters.
Given an initial guess with initial conditions $\bq_0$, period $T$, and swimming speed $v_0$, a periodic orbit can be obtained as a solution of the system of equations 
\begin{align}\label{eq:objective}
&{\bf g}(\Phi^{T}(\bq_0;v_0) - \bq_0)= {\bf 0},\,\,\,{\rm with} \\ \nonumber
&{\bf g}(\Delta \bq) = \left(\Delta \br, \sin\frac{\Delta \theta}{2}\right),
\end{align}
where $\Phi^{T}(\bq_0;v_0)$ is the time-$T$ flow map of Eqs.~\eqref{eq:model_vortex} with initial conditions $\bq_0$ and the parameter $v_0$ (we assume $\alpha$ is fixed throughout).
In other words, $\Phi^{T}(\bq_0;v_0) = \bq(T)$ with the initial condition $\bq(0) = \bq_0$ and swimming speed $v_0$.
The condition $\bf g = 0$ implies the conditions $\br(T) - \br_0 = {\bf 0}$ and $\theta(T) - \theta_0 = 2 \pi n$, for an integer $n$.
Fixing one or several of the initial guess parameters (for instance, $v_0$), a root-finding algorithm is used to adjust the remaining parameters in order to obtain a solution of Eq.~\eqref{eq:objective}.
If this procedure fails to converge to a solution, then either no periodic orbit with those fixed parameters exists, or the initial guess was not close enough to the true periodic orbit. 
For the root-finding algorithm, we use Matlab's \texttt{fsolve}.
Throughout the paper, we compute periodic orbits at discrete values of $\alpha \in [-1,0.97]$, beginning at $\alpha=-1$ and spaced at regular intervals $\Delta\alpha_p = 0.01$.
For values of $\alpha \geq 0.98$, our algorithm does not converge to the $t$-symmetric periodic orbit for $v_0 = 0.01$.
However, we checked that for $\alpha = 0.98$ and $v_0 < 0.01$ and sufficiently small, our algorithm is able to converge to the $t$-symmetric center-type orbit.
Thus, we conclude the orbit does not exist for $\alpha \geq 0.98$ and $v_0 \geq 0.01$, and we restrict our analysis to $\alpha \leq 0.97$.

For the calculation of the periodic orbits plotted in Fig.~\ref{fig:main_y0} at $v_0 =0.01$, our initial guesses for $\bq_0$ are given by Eq.~\eqref{eq:guess}, with $r^*$ (which depends on $\alpha$) given in Eq.~\eqref{eq:equilibrium_approx}.
Furthermore, we take the initial guess for the period as $T = 1$.
At each $\alpha$, we fix the parameters $v_0 = 0.01$, $x_0 = 0.25$, and $\theta_0 = 0$, so that we are guaranteed to locate a $t$-symmetric orbit, and then we allow the root-finding algorithm to adjust $y_0$ and $T$ to determine the periodic orbit.
Having obtained the exact periodic orbit at a fixed $v_0$, we can now increase $v_0$ by a small increment $\Delta v_{0,p}$ and use the previously calculated periodic orbit(s) to devise a good initial guess  $(y_0,T)$ at the new $v_0$.
This procedure is iterated until a step is reached at which the algorithm fails to converge to a periodic orbit.
In that case, we try to repeat the step with a smaller $\Delta v_{0,p}$, namely taking  $\Delta v_{0,p} \mapsto \Delta v_{0,p}/2$.
If this succeeds, we carry on with the new $\Delta v_{0,p}$.
If not, we continue to halve $\Delta v_{0,p}$ until a periodic orbit is successfully found or $\Delta v_{0,p}$ drops below a threshold value $\Delta v_{0,p} < 10^{-5}$.
Initial guesses for $y_0$ and $T$ at each step are obtained by quadratic extrapolation of $y_0$ and $T$ as a function of $v_0$ using the periodic orbits of the previous three steps. \cite{Fox2013}
 
Using the scheme described above, we are able to continue the main periodic orbit at each $\alpha$ shown in Fig.~\ref{fig:main_y0} with $v_0$ increasing from $v_0 = 0.01$,  until a critical $v_0$ at which our algorithm fails to converge to a periodic orbit.
Also, for a wide range of $\alpha < 0$, this orbit undergoes changes to its linear stability.
We show that both of these observations are due to bifurcations involving other periodic orbits.
We use variants of the previously described scheme to compute and continue these additional periodic orbits.
Here we describe one particularly important variant, which is the continuation of periodic orbits in $T$ rather than $v_0$.
The algorithm is basically the same, except the roles of $v_0$ and $T$ are swapped.
The procedure above is modified such that at each step, $T$ is held fixed while the root-finding algorithm adjusts the initial guesses $y_0$ and $v_0$ for the initial condition and swimming speed for which there exists a periodic orbit with fixed period $T$.
Then, $T$ is gradually incremented (in place of $v_0$) in order to continue the periodic orbit to higher periods.
As we show in Sec.~\ref{sec:bif}, this method is particularly effective at detecting saddle-node bifurcations (see Fig.~\ref{fig:saddlenode}).

We also supply the root-finding algorithm with explicitly calculated partial derivatives of Eq.~\eqref{eq:objective}, which improve the accuracy and performance of the algorithm.
The crux of this  is the partial derivatives of $\Phi^T$ with respect to $T$, $\bq_0$, and $v_0$.
The first of these is simply the phase-space velocity $\dot{\bq}$ evaluated at the final phase space point $\bq(T)$, i.e.\
\begin{equation*}
\frac{\partial \Phi^T(\bq_0;v_0)}{\partial T} = \dot{\bq}(\bq(T);v_0).
\end{equation*}
Defining the Jacobian matrix as the derivative of $\Phi^T$ with respect to the initial conditions $J(T) \equiv \partial \Phi^T/\partial \bq_0$, $J$ satisfies the differential equation \cite{ChaosBook}
\begin{equation}\label{eq:tangentflow}
\dot{J}(t) = A(\bq (t))J(t),
\end{equation}
with the initial condition $J(0) = {\rm Id}_{3\times3}$ and $A = \partial \dot{\bq} / \partial \bq$ as before.
Lastly, defining the derivative of $\Phi^T$ with respect to the parameter $v_0$ as ${\bf j}(T) \equiv \partial \Phi^T/ \partial v_0$, ${\bf j}$ satisfies the differential equation \cite{Gronwall1919}
\begin{equation}\label{eq:paramflow}
\dot{\bf j}(t) = A(\bq(t)) {\bf j}(t) + \frac{ \partial \dot{\bq}}{\partial v_0}(\bq(t); v_0),
\end{equation}
with the initial condition ${\bf j}(0) = {\bf 0}$.
Equations \eqref{eq:tangentflow} and \eqref{eq:paramflow} are numerically integrated along the trajectory $\bq(t)$ from $t= 0$ to $t = T$ in order to obtain $J(T)$ and ${\bf j}(T)$, respectively.
When a periodic orbit is found, the eigenvalues of $J(T)$ evaluated along the orbit determine the orbit's linear stability.

\subsubsection{Bifurcation analysis at a fixed $\alpha$}\label{sec:bif}
\begin{figure}
\centering
\includegraphics[width=0.6\textwidth]{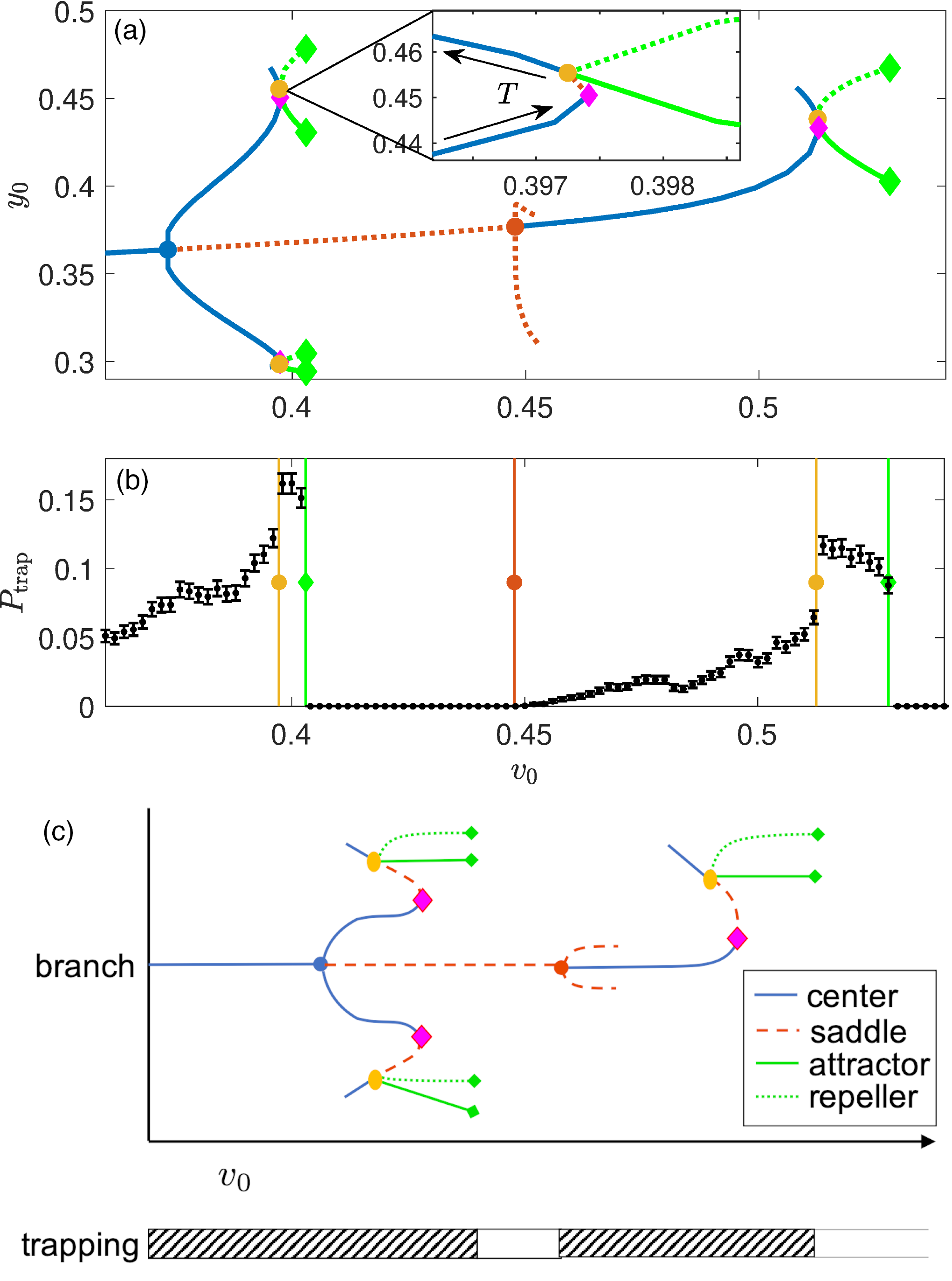}
\caption{Bifurcation diagram of the main periodic orbit for $\alpha$ near $-1$. (a) Initial condition $y_0$, with $\theta_0 = -\pi/4$, of numerically computed periodic orbits as a function of $v_0$ at $\alpha = -0.91$. Inset: magnification of a saddle-node bifurcation. Arrows indicate the direction of increasing orbit period $T$ along the concatenated lower and upper branches of periodic orbits. (b) Trapping probability $P_{\rm trap}$ as a function of $v_0$ for $\alpha=-0.91$. Error bars for nonzero values indicate $95\%$ confidence intervals. Vertical lines correspond to certain bifurcations in panel (a), see text. (c) Schematic illustrating the sequence of bifurcations undergone by the main periodic orbit for $-0.99 \leq \alpha \leq -0.82$ as $v_0$ increases. Each curve corresponds to a different family of periodic orbits, i.e.\ a branch. The dots represent bifurcations where additional periodic orbits are created as $v_0$ increases, while the diamonds represent bifurcations in which periodic orbits are destroyed. The legend shows the correspondence of line style and color to the periodic orbit's stability type, and is consistent with the styles in panel (a). The shaded regions in the rectangle below the orbit branches indicate ranges of $v_0$ where swimmer trapping inside a vortex cell can occur.}\label{fig:bifurcation_schematic}
\end{figure} 
Using the periodic orbit continuation schemes described in the previous section, we investigate the bifurcations undergone by the $t$-symmetric center periodic orbit as $v_0$ is increased.
Our approach consists of first identifying the sequence of bifurcations leading to the main orbit's changes of stability and eventual destruction at a fixed value of $\alpha$.
Indeed, the numerical signatures of a bifurcation are either changes in the linear stability of the orbit or a failure to continue the orbit past a fixed $v_0^*$.
Candidate bifurcation scenarios are hypothesized by examining Poincar\'{e} sections near bifurcations and studying properties of the periodic orbit as a bifurcation is approached.
Local bifurcation scenarios,  i.e.\ bifurcations leading to the creation or destruction of additional periodic orbits, are confirmed through the explicit numerical computation and continuation of the additional periodic orbits implicated in the bifurcation.
In this system we also find a particular type of global bifurcation known as a heteroclinic bifurcation, in which  the periodic orbit collides with multiple distinct swimming fixed points.
These are confirmed at a fixed $\alpha$ by looking for two signatures, namely the gradual approach of the orbit towards swimming fixed points and the divergence of the orbit period as the bifurcation is approached. 
Then, each bifurcation scenario is extended to nearby values of $\alpha$ through the automated computation of the additional periodic orbits implicated in the bifurcation.
In this section, we focus on the bifurcations at $\alpha = -0.91$, which is representative of the sequence of bifurcations applying to the range $-0.99 \leq \alpha \leq -0.82$.

\begin{figure}
\centering
\includegraphics[width=0.8\textwidth]{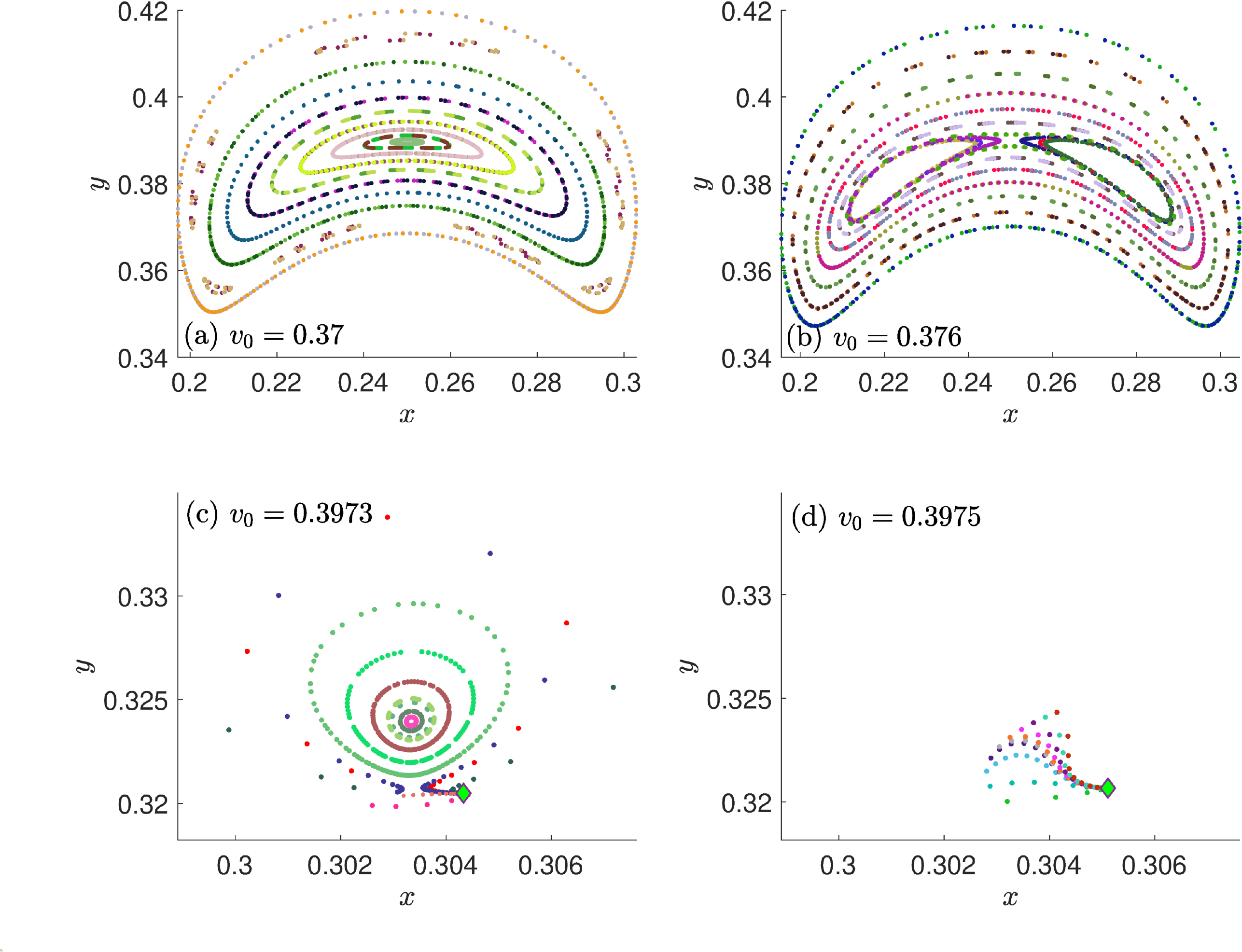}
\caption{Poincar\'{e} sections ($\theta=0,\,\,\dot{\theta} > 0$) illustrating certain bifurcations at $\alpha = -0.91$. (a) Near the main periodic orbit, before the supercritical pitchfork bifurcation, $v_0 = 0.37$. (b) Same region as panel (a), after the supercritical bifurcation, $v_0 = 0.376$. (c) Near one of the pitchfork center orbits, before the first saddle-node bifurcation, $v_0 = 0.3973$.  The green diamond is an attractor. (d) Same region as panel (c), after the first saddle-node bifurcation, $v_0 = 0.3975$. The green diamond is an attractor.}\label{fig:poincare_suppfk}
\end{figure}

Figure~\ref{fig:bifurcation_schematic} shows the sequence of bifurcations at $\alpha = -0.91$, with numerically computed periodic orbit initial conditions in  Fig.~\ref{fig:bifurcation_schematic}a, the corresponding trapping probability $P_{\rm trap}$ in Fig.~\ref{fig:bifurcation_schematic}b (taken from the calculation presented in Fig.~\ref{fig:monte_carlo}), and a schematic illustration in Fig.~\ref{fig:bifurcation_schematic}c.
Here, we plot the initial condition $y_0$ under the constraint $\theta_0 = -\pi/4$.
When the periodic orbit is $y=x$ $t$-symmetric, as is the case for the main periodic orbit, then the additional constraint on the initial conditions $x_0 = y_0$ is satisfied (see Table \ref{tab:tsymm}).
The main orbit remains of center stability type until $v_0 \approx 0.37$, at which point the orbit transitions to a $t$-symmetric saddle through a supercritical pitchfork bifurcation, in which two additional $t$-symmetric center periodic orbits are created.
We refer to these orbits as pitchfork centers, to distinguish them from the main periodic orbit.
This scenario is clearly evident on the Poincar\'{e} sections in Figs.~\ref{fig:poincare_suppfk}a and \ref{fig:poincare_suppfk}b, showing the region around the main periodic orbit before and after the bifurcation.
Note that the choice $\theta=0$ for the Poincar\'{e} section means that the initial conditions plotted in Fig.~\ref{fig:bifurcation_schematic}a (with $\theta_0 = -\pi/4$) do not correspond to the locations of the periodic orbits on this surface of section.
After the bifurcation, the elliptical region in the immediate vicinity of the main periodic orbit splits into two distinct elliptical regions (Fig.~\ref{fig:poincare_suppfk}b), each surrounding one of the newly created pitchfork centers.
The new orbits break the $\pi/2$ rotational symmetry of the main periodic orbit.
However, they are invariant under rotations by $\pi$ about the vortex center, and each of the pitchfork centers maps into the other by the $\pi/2$ rotational symmetry. 
As seen in Fig.~\ref{fig:poincare_suppfk}b, the pitchfork centers also break the vertical-axis $t$-symmetry because they do not lie on the $x=0.25$ axis; however, they are still $y=x$ $t$-symmetric.
Hence, we select initial conditions satisfying the constraints  $y_0 = x_0$ and $\theta_0 = -\pi/4$ in the continuation computations for these orbits, which apply as well to the initial conditions plotted in Fig.~\ref{fig:bifurcation_schematic}a.

After the supercritical pitchfork bifurcation, the existence of the pitchfork centers guarantees a nonzero probability of swimmer trapping as $v_0$ is increased past this bifurcation, as is confirmed by Fig.~\ref{fig:bifurcation_schematic}b.
Following the branches of the newly created center orbits as $v_0$ is increased further, Fig.~\ref{fig:bifurcation_schematic} shows that each orbit is eventually destroyed in a saddle-node bifurcation, i.e.\ a collision with a saddle from a secondary branch of periodic orbits.
This situation for one of the orbits is magnified in the inset of Fig.~\ref{fig:bifurcation_schematic}a and seen clearly in Fig.~\ref{fig:bifurcation_schematic}c.
We also show Poincar\'{e} sections before and after the saddle-node bifurcation near one of the pitchfork centers in Figs.~\ref{fig:poincare_suppfk}c and \ref{fig:poincare_suppfk}d, respectively.
Before the bifurcation, we see the pitchfork center, surrounded by invariant tori, and there is a hint of a saddle below the tori from the orbits tracing out hyperbola-shaped curves.
As the orbits near the saddle are mapped forward in time, they appear to accumulate in the region near $(x,y) = (0.304,0.32)$.
Indeed, we find that there is an attractor [specifically, an attracting limit cycle of Eqs.~\eqref{eq:model_vortex}] near this point, indicated by the green diamond in Fig.~\ref{fig:poincare_suppfk}c.
As $v_0$ is increased past the bifurcation, the saddle and pitchfork center collide and disappear completely, while the attractor persists (Fig.~\ref{fig:poincare_suppfk}d) and continues to trap swimmers (Fig.~\ref{fig:bifurcation_schematic}b).

\begin{figure}
\centering
\includegraphics[width=0.6\textwidth]{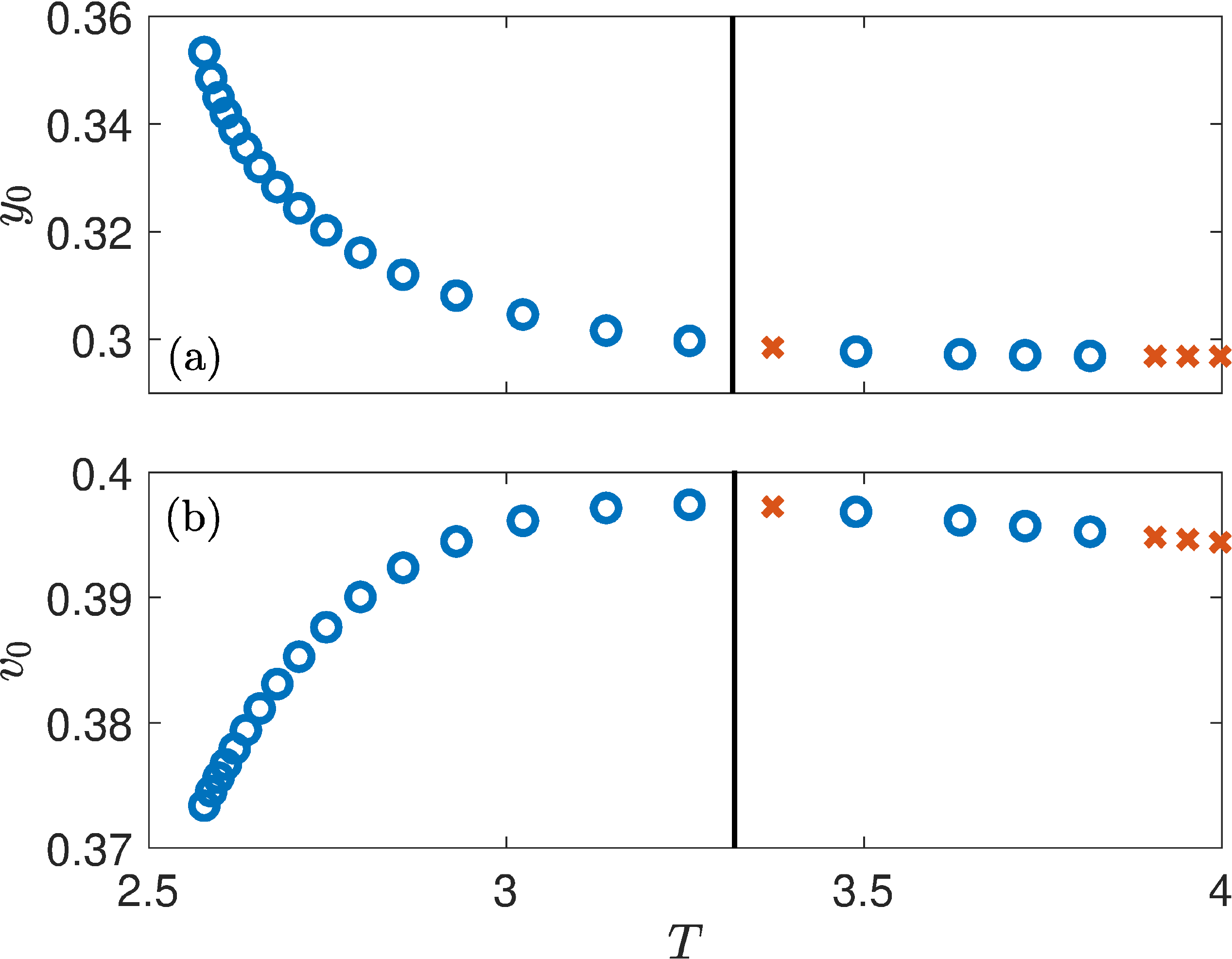}
\caption{Numerical continuation of the pitchfork center orbit with increasing $T$ at $\alpha = -0.91$. (a) Initial condition $y_0$, with $x_0 = y_0$ and $\theta_0 = -\pi/4$, as a function of $T$. (b) Swimming speed $v_0$ for which the periodic orbit with period $T$ exists. In both panels, blue circles indicate centers, while orange exes indicate saddles. The black line indicates the transition from the pitchfork center branch, on the left of the line, to the secondary periodic orbit branch, on the right.}\label{fig:saddlenode}
\end{figure}
By continuing the saddle and attractor to lower values of $v_0$, we find that they originate from the same $t$-symmetry breaking bifurcation of a secondary branch of $t$-symmetric center orbits.
The saddle is $y=x$ $t$-symmetric, like the pitchfork center (in fact, it can be shown that it must have the same symmetry as the pitchfork center using an argument based on the Poincar\'{e} index).\cite{Roberts1992}
Furthermore, we observe that the period $T$ increases monotonically as one moves along the continuous curve in parameter space $(y_0,v_0)$ obtained by concatenating the branch of pitchfork centers with the secondary branch of orbits including the saddles, as indicated in the inset of Fig.~\ref{fig:bifurcation_schematic}a, in the direction pitchfork center $\rightarrow$ saddle.
Taking advantage of this observation, we numerically continue the pitchfork centers through the saddle-node bifurcation and subsequently continue the secondary branch of orbits towards lower $v_0$ with a single computation, by continuing the orbits in $T$ rather than $v_0$, as described in Sec.~\ref{sec:continuation}.
The results of this calculation for $\alpha = -0.91$ are shown in Fig.~\ref{fig:saddlenode}, where each point is a numerically computed periodic orbit.
We see that $v_0$ attains a maximum as $T$ is increased, and around this point the stability of the found orbits changes from center to saddle.
This is a signature of a saddle-node bifurcation: it suggests that as $v_0$ approaches this maximum from below, there are two periodic orbits with initial conditions $y_0$ approaching each other, the pitchfork center on the left of the maximum and the secondary orbit with saddle stability type on the right.
We confirmed this scenario by refining the calculation leading to Fig.~\ref{fig:saddlenode} by decreasing the steps $\Delta T$ in the vicinity of the stability change near the maximum of $v_0$.

Note that if the steps $\Delta T$ taken are too large, one might miss the transition to saddle stability-type entirely, since the secondary orbit quickly changes to a center as $T$ is increased further past the maximum in $v_0$.
Equivalently, beginning on the center-stability part of the secondary orbit branch, we see that the secondary orbit becomes a saddle as $v_0$ is increased.
We identify this transition from center to saddle as a $t$-symmetry breaking bifurcation that simultaneously creates an additional pair of asymmetric orbits: an attractor and a repeller. \cite{Politi86,Roberts1992}
The attractor of this pair is in fact the one we see in Figs.~\ref{fig:poincare_suppfk}c and \ref{fig:poincare_suppfk}d.
The attractor-repeller pairs of orbits are plotted as the green curves in Fig.~\ref{fig:bifurcation_schematic}.
This pair of orbits breaks the $y=x$ $t$-symmetry of the upper branch of secondary orbits, meaning that with the initial condition $\theta_0 = -\pi/4$ fixed, neither the attractor's $(x_0,y_0)$ nor the repeller's $(x_0,y_0)$ are on the $t$-symmetry axis, i.e.\ $x_0 \neq y_0$.
The orbits are linked to each other by the broken $t$-symmetry.
In particular, the attractor and repeller's initial conditions map into one another through reflection about the $y=x$ axis.
Hence, in our periodic orbit computations, we only compute the attractor and obtain the initial conditions of the repeller by symmetry.
The lower pair of attractor-repeller orbits and the lower secondary branch of $t$-symmetric orbits seen in Fig.~\ref{fig:bifurcation_schematic} are related to the corresponding upper set of orbits by the $\pi/2$-rotational symmetry about the vortex center.

\begin{figure}
\centering
\includegraphics[width=0.8\textwidth]{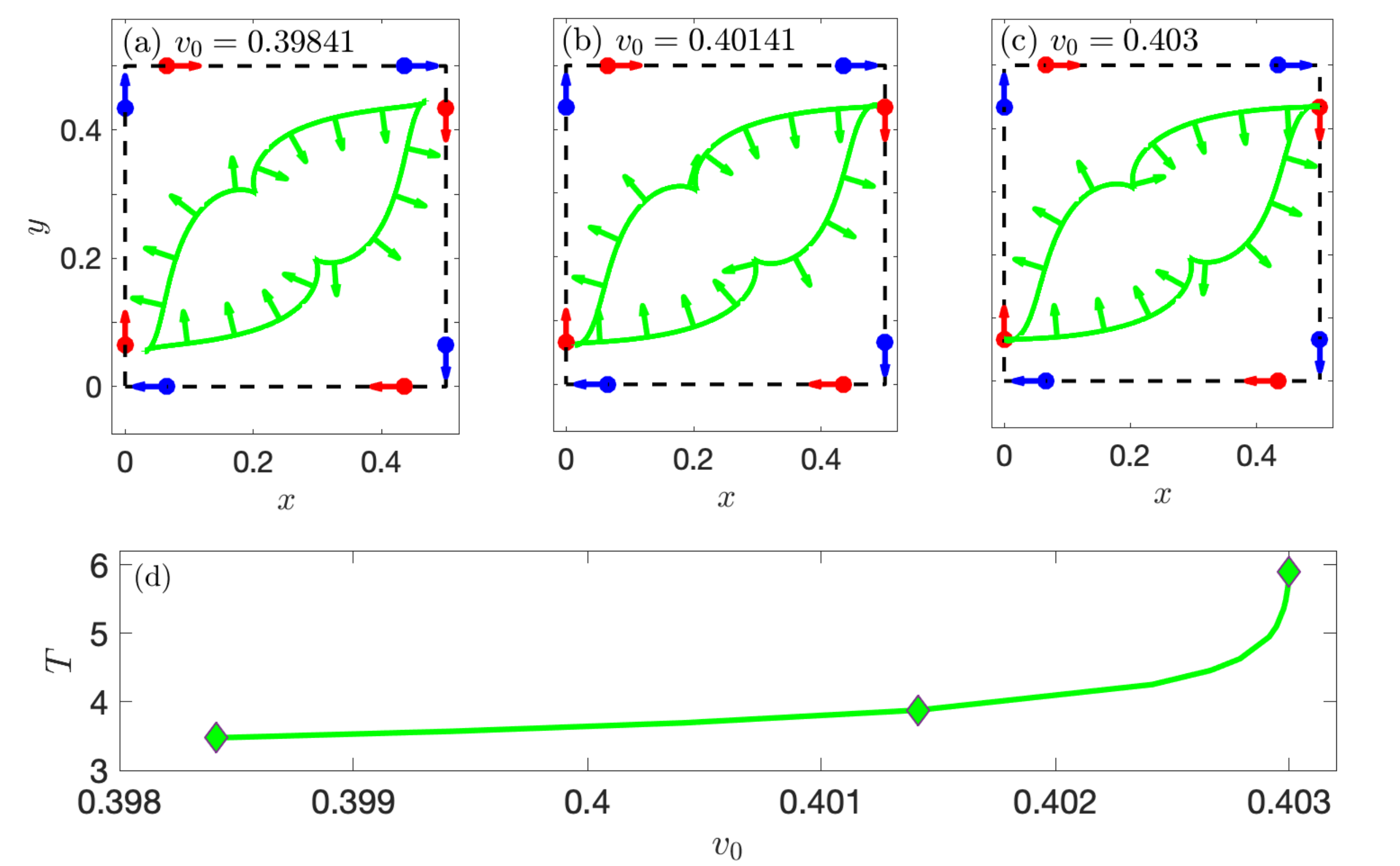}
\caption{Heteroclinic bifurcation of an attractor associated with the pitchfork centers at $\alpha=-0.91$. The attractor is plotted as a green curve for $v_0 = 0.39841$ (a), $v_0 = 0.40141$ (b), and $v_0 = 0.403$ (c). In each of these panels, red dots are SSU swimming fixed points, while blue dots are SUU swimming fixed points, and arrows indicate the swimmer orientation. Dashed lines indicate the boundary of the vortex cell. Swimmers traverse the orbits counterclockwise. (d) Attractor period $T$ as a function of $v_0$. Diamonds indicate $v_0$ values for which the attractor is plotted in panels (a)--(c).}\label{fig:attractor_bif1}
\end{figure}
The attractor-repeller pairs of orbits persist as $v_0$ is increased until they are destroyed in heteroclinic bifurcations consisting of collisions with swimming fixed points on the vortex cell boundary, as illustrated in Fig.~\ref{fig:attractor_bif1} for $\alpha = -0.91$.
Figures \ref{fig:attractor_bif1}a--c show the upper attractor orbit from Fig.~\ref{fig:bifurcation_schematic} at three values of $v_0$ as the bifurcation is approached.
The leftmost and rightmost points on the orbit get closer and closer to the SSU swimming fixed points on the left and right vortex cell boundaries, respectively, as $v_0$ increases.
The swimming direction at the extremal points of the orbit also approaches the SSU swimming fixed points' swimming direction ($+\hat{\bf y}$ for the left fixed point and $-\hat{\bf y}$ for the right fixed point), implying that these points on the attractor approach the swimming fixed points in the full phase space.
Meanwhile, the period $T$ of the attractor rapidly increases as a critical value of $v_0$ is approached, as shown in Fig.~\ref{fig:attractor_bif1}d.
This behavior is consistent with the heteroclinic bifurcation scenario, in which $T$ should diverge as the bifurcation is approached because the swimmer spends more and more time near the swimming fixed points as it gets closer to them.\cite{Strogatz}
Clearly, this scenario also applies to the repeller obtained by reflecting the attractor plotted in Figs.~\ref{fig:attractor_bif1}a--c about the $y=x$ axis, because the repeller collides with the SUU swimming fixed points on the upper and lower vortex cell boundaries.
Likewise, it applies to the lower attractor-repeller pair from Fig.~\ref{fig:bifurcation_schematic} by the $\pi/2$-rotational symmetry.

As $v_0$ is increased past the heteroclinic bifurcation, the only periodic orbit remaining in our analysis is the main $t$-symmetric orbit, which is a saddle and thus unstable.
This suggests that there may be no way for swimmers to remain trapped inside a vortex cell for these swimming speeds, as indicated by the interruption in the shaded bar in Fig.~\ref{fig:bifurcation_schematic}c.
In fact, this prediction is borne out by the numerical results shown in Fig.~\ref{fig:bifurcation_schematic}b, where $P_{\rm trap}$ abruptly drops to zero as $v_0$ is increased past the first vertical green line marking the heteroclinic bifurcation.
We see in Fig.~\ref{fig:bifurcation_schematic}a that the next bifurcation occurs at $v_0 \approx 0.45$ (when $\alpha = -0.91$), when the main orbit transitions from saddle to center.
This occurs via a subcritical pitchfork bifurcation, meaning two additional $t$-symmetric saddles which break the $\pi/2$ rotational symmetry are created in this bifurcation. 
We find that these saddles are $t$-symmetric with respect to the horizontal and vertical axes through the vortex center, but not the diagonal axes (i.e.\ with $\theta_0 = -\pi/4$, we have $x_0 \neq y_0$).
Hence, swimmer trapping again becomes possible in the vicinity of the stable main periodic orbit after the subcritical pitchfork bifurcation.
This is also confirmed by Fig.~\ref{fig:bifurcation_schematic}b, where we see $P_{\rm trap}$ begins increasing after $v_0$ exceeds the orange line marking the subcritical pitchfork bifurcation.

\begin{figure}
\centering
\includegraphics[width=0.8\textwidth]{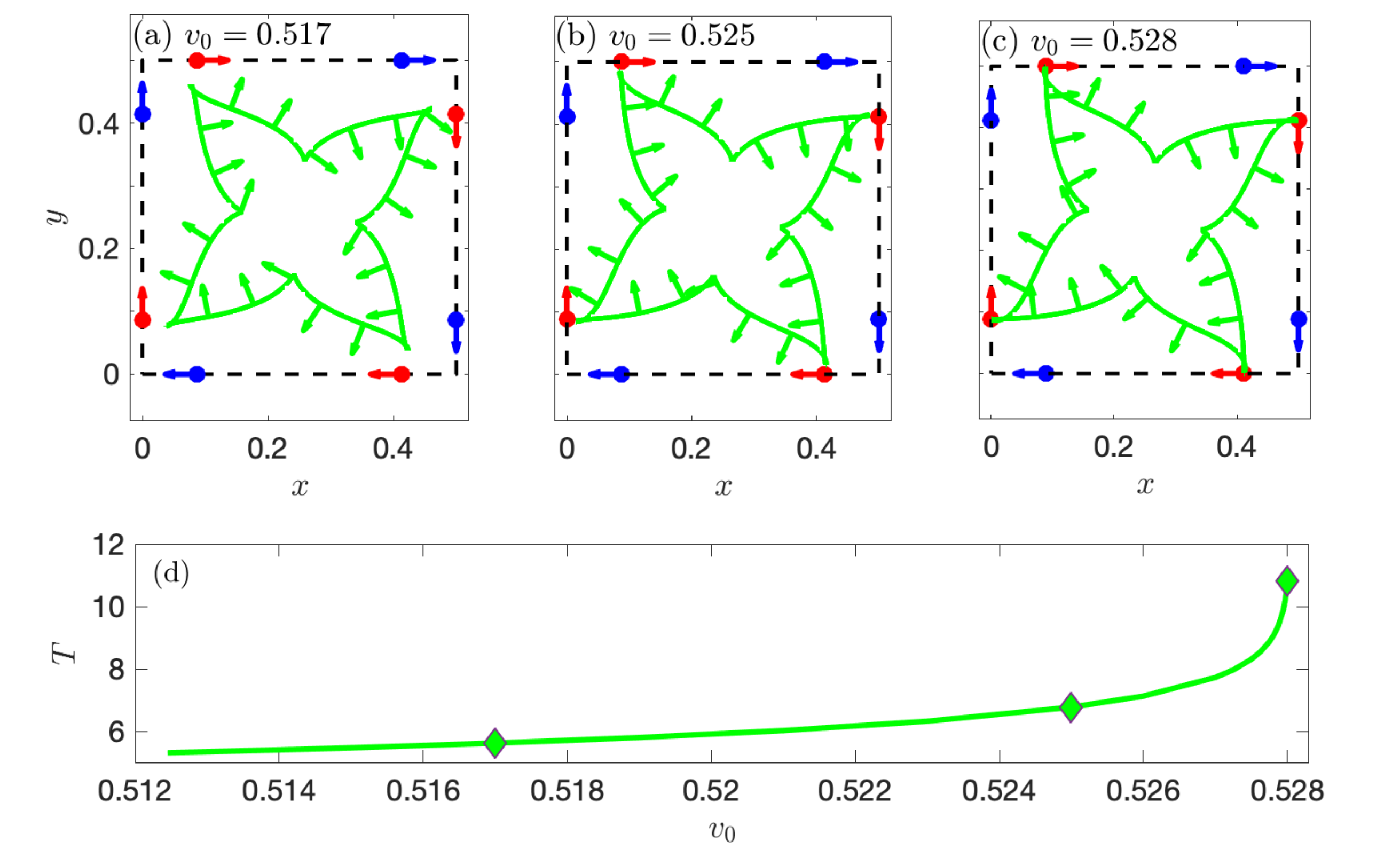}
\caption{Heteroclinic bifurcation of the attractor associated with the main orbit at $\alpha=-0.91$. The attractor is plotted as a green curve for $v_0 = 0.517$ (a), $v_0 = 0.525$ (b), and $v_0 = 0.528$ (c).  In each of these panels, red dots are SSU swimming fixed points, while blue dots are SUU swimming fixed points, and arrows indicate the swimmer orientation. Dashed lines indicate the boundary of the vortex cell. Swimmers traverse the orbits counterclockwise. (d) Attractor period $T$ as a function of $v_0$. Diamonds indicate $v_0$ values for which the attractor is plotted in panels (a)--(c).}\label{fig:attractor_bif2}
\end{figure}
The final sequence of bifurcations in which the main periodic orbit is destroyed is qualitatively the same as the sequence leading to the destruction of the pitchfork centers.
This is seen most clearly in Fig.~\ref{fig:bifurcation_schematic}c: the main orbit collides with a secondary $t$-symmetric saddle in a saddle-node bifurcation.
We again find that the orbit period $T$ increases monotonically as one follows the main $t$-symmetric orbit branch through the saddle-node bifurcation onto the secondary branch of orbits.
The secondary saddle is created through the same type of $t$-symmetry breaking bifurcation (center $\rightarrow$ saddle, attractor, repeller) that we observe along the secondary branches associated with the pitchfork centers.
Lastly, the asymmetric attractor-repeller pair born out of this bifurcation persists as $v_0$ is increased past the saddle-node bifurcation of the main orbit, and the pair is itself destroyed through a heteroclinic bifurcation with the swimming fixed points.
This bifurcation is depicted in Fig.~\ref{fig:attractor_bif2}.
We again see the signatures of a heteroclinic bifurcation: a diverging period (Fig.~\ref{fig:attractor_bif2}d), and the approach of extremal points on the orbit towards swimming fixed points in the full $(\br,\theta)$ phase space (Figs.~\ref{fig:attractor_bif2}a--c). 
In this case, because the attractor is invariant under the $\pi/2$ rotational symmetry, it approaches all the SSU swimming fixed points on the boundary of the vortex cell simultaneously.
After this last bifurcation, there are no periodic orbits remaining in our analysis at this value of $\alpha$.
Correspondingly, we see that swimmer trapping ceases for higher values of $v_0$ in Fig.~\ref{fig:bifurcation_schematic}b, where the last green line is the heteroclinic bifurcation of the orbit plotted in Fig.~\ref{fig:attractor_bif2}.

To summarize, we have fully characterized the sequence of bifurcations undergone by the main periodic orbit at $\alpha = -0.91$.
We have shown that this allows the accurate prediction of the whether or not swimmer trapping is possible, i.e.\ whether $P_{\rm trap} > 0$ or $P_{\rm trap} = 0$, at a given value of $v_0$.
Essentially, if a stable periodic orbit (either a center or attractor) confined to a vortex cell exists, then swimmer trapping is possible; otherwise, all swimmers eventually escape their initial vortex cell.
In particular, our analysis explains the counter-intuitive breakdown and reemergence of trapping for intermediate swimming speeds.
Trapping breaks down following the heteroclinic bifurcation of the attractors associated with the pitchfork centers; at higher $v_0$, it reemerges following the subcritical pitchfork bifurcation of the main periodic orbit.
In the next section, we extend these results to all values of $\alpha$, i.e.\ for all swimmer shapes and both perpendicular and parallel swimmers.
\subsubsection{Bifurcations for all swimmer shapes}
\begin{figure}
\centering
\includegraphics[width=0.9\textwidth]{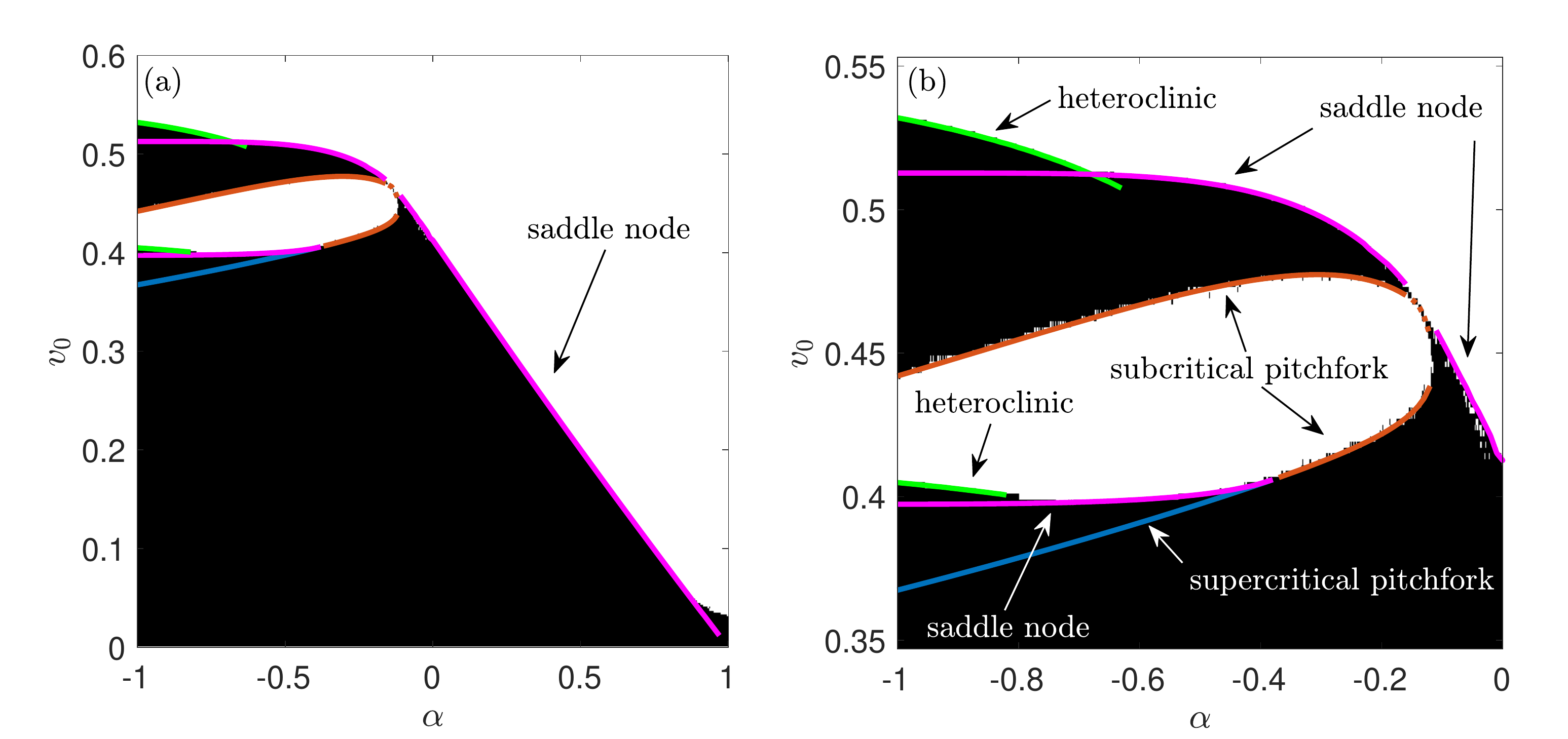}
\caption{Trapping and periodic orbit bifurcations. Black regions indicate parameter regions where trapping occurs, while white regions indicate regions where no trapping occurs (same as Fig.~\ref{fig:monte_carlo}b). Labeled, colored curves indicate various periodic orbit bifurcations (see text). (a) Full range of parameters $(v_0,\alpha)$. (b) Magnification of the intermediate $v_0$ region for perpendicular swimmers, where trapping ceases, reemerges, and ceases again as $v_0$ increases.}\label{fig:phase_diagram}
\end{figure}
We continue the families of periodic orbits associated with bifurcations of the main periodic orbit for $\alpha = -0.91$ at all values of $\alpha$ considered in our analysis, with the objective of identifying the bifurcation scenarios associated with the breakdown of swimmer trapping for all swimmer shapes and relative swimming directions.
The results of these calculations are summarized in Fig.~\ref{fig:phase_diagram}, where we show periodic orbit bifurcation curves as well as parameter regions of zero and nonzero trapping, taken from Fig.~\ref{fig:monte_carlo}b.
We have determined that the sequence of bifurcations with increasing $v_0$ schematically illustrated in Fig.~\ref{fig:bifurcation_schematic}c applies for all $-0.99 \leq \alpha \leq -0.82$.
In particular, the sequence of bifurcations leading to the first breakdown of trapping comprises the supercritical pitchfork bifurcation of the main periodic orbit (lower blue curve in Fig.~\ref{fig:phase_diagram}), the saddle-node bifurcation of the pitchfork centers (lower magenta curve in Fig.~\ref{fig:phase_diagram}b), and finally the heteroclinic bifurcation of the attractors associated with the pitchfork centers (lower green curve).
We see clearly in Fig~\ref{fig:phase_diagram}b that the heteroclinic bifurcation curve is consistent with the values of $v_0$ above which trapping ceases  for this range of $\alpha$.
Then, trapping reemerges due to a subcritical pitchfork bifurcation of the main periodic orbit (upper orange curve).
Finally, trapping ceases again due to a saddle-node bifurcation of the main periodic orbit with a secondary saddle born out of a $t$-symmetry breaking bifurcation (upper magenta curve in Fig.~\ref{fig:phase_diagram}b), for which the corresponding attractor persists to even higher values of $v_0$ until it is destroyed in a heteroclinic bifurcation (upper green curve).

\begin{figure}
\includegraphics[width=0.7\textwidth]{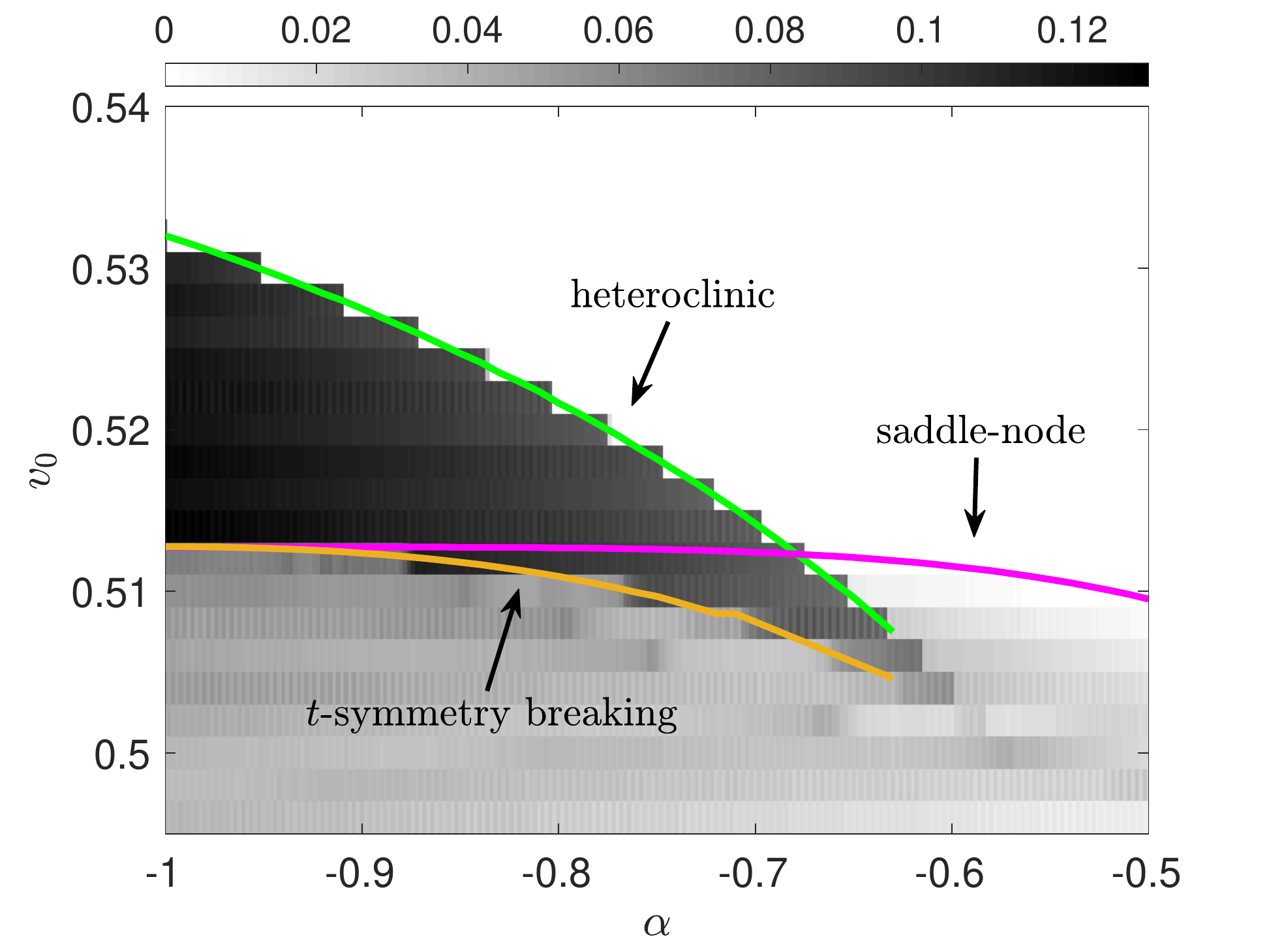}
\caption{Bifurcations and jumps in trapping probability. The trapping probability $P_{\rm trap}$ at each $(v_0,\alpha)$ is indicated by the gray scale. The labeled colored curves are bifurcation curves corresponding to the $t$-symmetry breaking bifurcation of the secondary orbit (yellow), the saddle-node bifurcation of the main orbit (magenta), and the heteroclinic bifurcation of the attractor (green).}\label{fig:tsymmbreak}
\end{figure}

For $\alpha = -1$, our periodic orbit computations suggest a slight modification of the sequence of bifurcations described above.
As $\alpha \rightarrow -1$, we find that the swimming speed $v_0^{\rm sn}$ at which the pitchfork center saddle-node bifurcation occurs approaches the swimming speed $v_0^{t}$ at which the $t$-symmetry breaking bifurcation occurs.
This corresponds to the secondary saddle branches shown in Fig.~\ref{fig:bifurcation_schematic}c getting shorter and shorter as $\alpha \rightarrow -1$.
At $\alpha = -1$, we do not find saddles for $v_0$ values close to the disappearance of the pitchfork centers, but we do find the secondary center orbits.
The confluence of the saddle-node and $t$-symmetry breaking bifurcation curves suggests a codimension-two bifurcation scenario, in which two $t$-symmetric centers collide and give rise to an attractor-repeller pair.
We also observe this phenomenon at $\alpha = -1$ near the disappearance of the main branch periodic orbit (see Fig.~\ref{fig:tsymmbreak}).
Aside from the saddle-node and $t$-symmetry breaking bifurcations, all the bifurcations associated with the main periodic orbit at $\alpha = -1$ are the same as described previously.

As $\alpha$ increases, the sequence of bifurcations leading to the breakdown of trapping undergoes qualitative changes as critical values of $\alpha$ are crossed.
The first of these changes concerns the heteroclinic bifurcations of the attractors.
In Fig.~\ref{fig:phase_diagram}b, we observe that the swimming speeds $v_0^{\rm h}$ at which the heteroclinic bifurcations occur decrease as $\alpha$ increases.
This is true of both the heteroclinic bifurcations associated with the pitchfork centers (lower green curve) and the one associated with the main periodic orbit (upper green curve).
There is then a value of $\alpha$ past which $v_0^{\rm h} < v_0^{\rm sn}$, i.e.\ the heteroclinic bifurcation occurs at a lower swimming speed than the corresponding saddle-node bifurcation.
For both families of heteroclinic bifurcations, this occurs at the intersection between the green and magenta curves in Fig.~\ref{fig:phase_diagram}b.
We see this clearly for the upper heteroclinic bifurcation curve $v_0^{\rm h}$ in Fig.~\ref{fig:phase_diagram}, which intersects the saddle-node bifurcation curve around $\alpha = -0.68$.
However, the transition to $v_0^{\rm h} < v_0^{\rm sn}$ occurs first for the attractors associated with the pitchfork centers, around $\alpha = -0.82$.
We did not numerically compute the heteroclinic bifurcation curve associated with these attractors past the point $v_0^{\rm h} < v_0^{\rm sn}$, though we expect we would also observe an intersection of $v_0^{\rm h}$ and $v_0^{\rm sn}$,  as seen for the main orbit.
For higher values of $\alpha$ past this intersection, the periodic orbit responsible for trapping in the range $v_0^{\rm h} < v_0 < v_0^{\rm sn}$ is the pitchfork center.
When this orbit is destroyed by the saddle-node bifurcation, trapping ceases.
This scenario is in agreement with the simulation results plotted in Fig.~\ref{fig:phase_diagram} for both $v_0^{\rm h} < v_0^{\rm sn}$ transitions, i.e.\ the one for the pitchfork centers occurring around $\alpha = -0.82$ and the one for the main orbit occurring around $\alpha = -0.68$.

The next qualitative change with increasing $\alpha$ occurs when the supercritical pitchfork bifurcation curve intersects the pitchfork centers' saddle-node bifurcation curve, around $\alpha = -0.37$.
For higher values of $\alpha$, the main periodic orbit no longer creates the pitchfork centers when it changes stability from center to saddle with increasing $v_0$.
Instead, the secondary saddles involved in the pitchfork centers' saddle-node bifurcation persist for $\alpha \geq -0.37$, and they collide with the main center periodic orbit when it changes stability.
Hence, for $\alpha \geq -0.37$, the main orbit goes unstable through a subcritical pitchfork bifurcation occurring at the swimming speed $v_0^{\rm sub}$, and subsequently trapping ceases, as seen in Fig.~\ref{fig:phase_diagram}b.
Note that in this case, the saddles exist below the bifurcation curve, i.e.\ for $v_0 < v_0^{\rm sub}$, and they are destroyed by the bifurcation.

\begin{figure}
\includegraphics[width=0.7\textwidth]{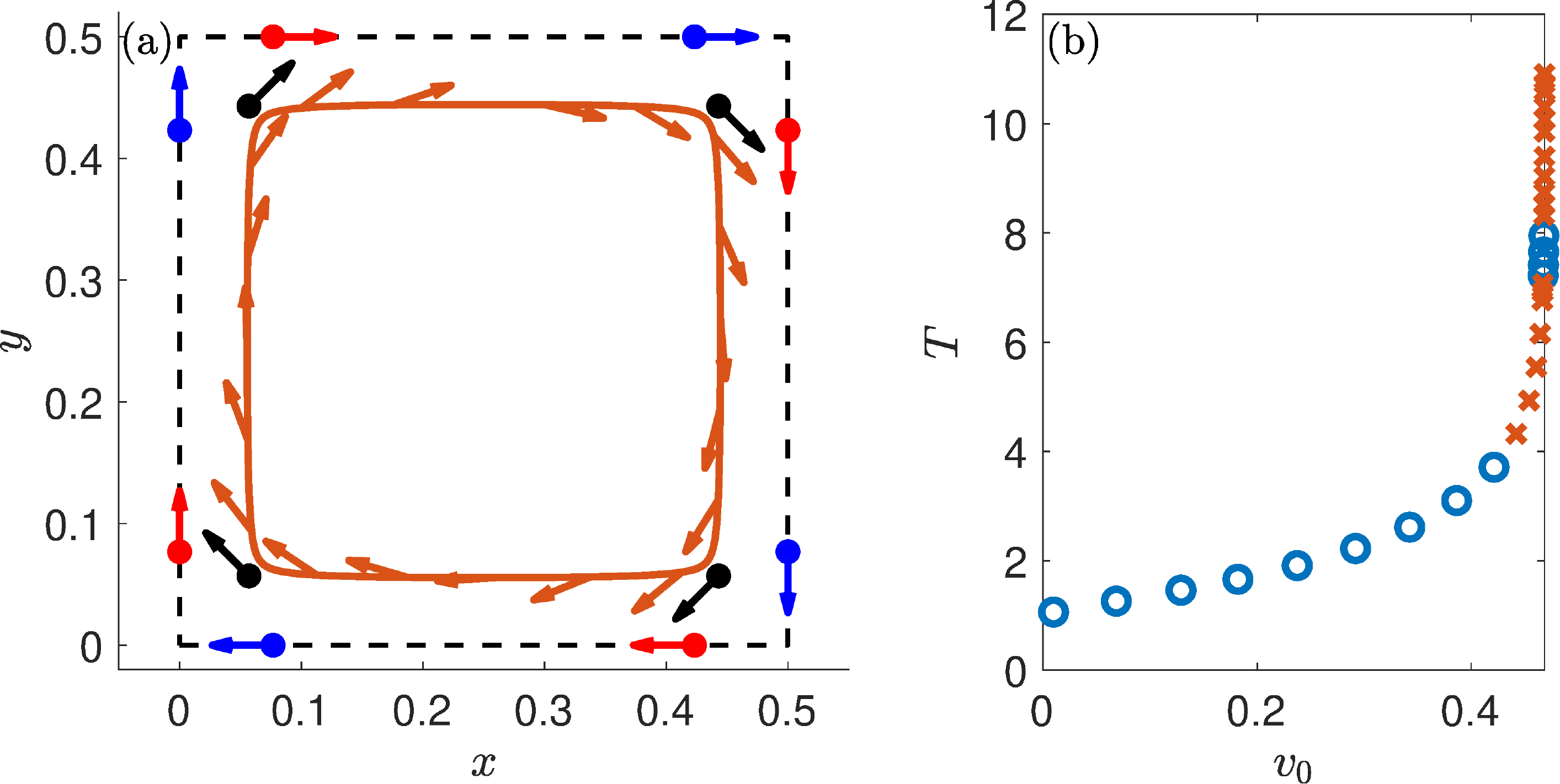}
\caption{Main periodic orbit near the final bifurcation for $\alpha = -0.14$. (a) Swimmer trajectory for $v_0 =  v_{\rm birth}(\alpha) = 0.46417$, where the orbit period $T = 6.1$ and the orbit is a saddle. The $t$-symmetric secondary swimming fixed points are plotted as the black dots. (b) Orbit period $T$ as a function of $v_0$, from numerical continuation computations with increasing $T$. Blue circles indicate centers, while orange exes indicate saddles.}\label{fig:main_orbit_sec_het}
\end{figure}
As $\alpha \rightarrow -0.15$, we observe in Fig.~\ref{fig:phase_diagram}b that the upper subcritical pitchfork bifurcation curve, corresponding to the main orbit regaining stability by transitioning from saddle to center, approaches the saddle-node bifurcation curve of the main orbit.
At the same time, in the range $-0.15 \leq \alpha \leq -0.12$, we find that the main orbit period $T$ grows rapidly over a small range of $v_0$, with $T > 10$ surpassed in the orbit $T$-continuation computations before a bifurcation causing the destruction of the main orbit is reached.
This is shown in Fig.~\ref{fig:main_orbit_sec_het}b for $\alpha = -0.14$, where we see also that the main orbit undergoes two transitions from center to saddle.
The first transition remains a subcritical pitchfork bifurcation, as described in the previous paragraph, while we have not investigated the bifurcation scenario of the second transition.
We have only observed the second center $\rightarrow$ saddle transition for $-0.14 \leq \alpha \leq -0.12$.
As $v_0$ increases beyond this center $\rightarrow$ saddle transition, the rapid growth of $T$ suggests a final heteroclinic bifurcation with swimming fixed points.
However, unlike the case of the attractors discussed in Sec.~\ref{sec:bif} (Figs.~\ref{fig:attractor_bif1} and \ref{fig:attractor_bif2}), we do not observe the swimmer trajectory approaching any of the primary swimming fixed points on the vortex cell boundary.
On the other hand, the rapid growth of $T$ occurs very close to the parameter values where the secondary swimming fixed points are born, i.e.\ close to the curve in parameter space $v_{\rm birth}(\alpha)$  given by Eq.~\eqref{eq:vbirth}  (see Fig.~\ref{fig:equilbria_stabilty}).
As shown in Fig.~\ref{fig:main_orbit_sec_het}a, where we have plotted the swimmer trajectory on the main periodic orbit for $\alpha = -0.14$ and $v_0 =  v_{\rm birth}(\alpha) = 0.46417$, the swimmer orbit does indeed contain points which get very close to the $t$-symmetric secondary swimming fixed points.
Hence, we speculate that in the vicinity of these parameters, there is a heteroclinic bifurcation consisting of the collision of the main orbit with the secondary swimming fixed points.
Due to the very narrow range of parameters over which these bifurcations occur, we have not determined the bifurcation scenarios to any more detail than discussed here.
For the range $-0.15 \leq \alpha \leq -0.12$, we have plotted the location of the main orbit's saddle $\rightarrow$ center transition (which is likely a subcritical pitchfork bifurcation, as for $\alpha \leq -0.16$) as the dotted orange curve in Fig.~\ref{fig:phase_diagram}b.
We observe that this curve is very close to the breakdown of trapping with increasing $v_0$.

Following this transition region, the bifurcation leading to the breakdown of trapping for $-0.11 \leq \alpha \leq 0.97$ is simply a single saddle-node bifurcation of the main orbit, plotted as the rightmost magenta curve in Fig.~\ref{fig:phase_diagram}.
We have not detected any changes in the stability of the main orbit up to the destruction of the orbit in the saddle-node bifurcation, in contrast to swimmers with $\alpha \leq -0.12$.
This is consistent with the persistence of swimmer trapping up to $v_0^{\rm sn}$ that we see in our simulations (Fig.~\ref{fig:phase_diagram}).
Furthermore, the breakdown of swimmer trapping for $v_0 > v_0^{\rm sn}$ for most values of $\alpha \geq -0.11$ suggests an absence of other stable periodic orbits that trap swimmers.
This stands in contrast to the case of $-1 \leq \alpha \leq -0.68$, where an attractor continues to trap swimmers for a range of $v_0 > v_0^{\rm sn}$.
Hence, we have not investigated in detail the bifurcation scenarios with decreasing $v_0$ of the saddle involved in the main orbit's saddle-node bifurcation.
We remark that the period and maximum eigenvalue of the saddle grow rapidly as $v_0$ decreases for a range of $\alpha$ values; hence we find $T$-continuation to be much more effective at locating and continuing this saddle than $v_0$-continuation.
In Fig.~\ref{fig:phase_diagram}a, we see that for $\alpha$ near $1$, the trapping probability does not vanish until some value of $v_0 > v_0^{\rm sn}$.
We expect that this is a finite-time effect, due to the slow escape of swimmers with small $v_0$ from the vortex center.
This is plausible because when $v_0 \rightarrow 0$, we expect that the swimmer dynamics approaches passive particle dynamics, and passive particles are confined to individual vortex cells for infinite time.

We note that Fig.~\ref{fig:phase_diagram}a suggests that $v_0^{\rm sn}$ has a linear dependence on $\alpha$ for $-0.11 \leq \alpha \leq 0.97$.
This is consistent with the heuristic argument for the linear dependence of $v_0^*$ on $\alpha$ for parallel swimmers $0 \leq \alpha \leq 1$ presented in Ref.~\onlinecite{Torney2007}, where $v_0^*$ is the critical swimming speed for the breakdown of trapping.
In particular, we expect that $v_0^{\rm sn} \rightarrow 0$ as $\alpha \rightarrow 1$, leading to the guaranteed eventual escape of all $\alpha = 1$ swimmers.
Our results suggest that the linear scaling also holds for nearly circular perpendicular swimmers (i.e. for $-0.11 \leq \alpha < 0$), but it breaks down for sufficiently elongated perpendicular swimmers.
The main periodic orbit of such swimmers undergoes a much richer sequence of bifurcations, with the details depending on how elongated the swimmer is.
In general, the bifurcations are such that trapping ceases and reemerges as the swimming speed increases, because the orbits responsible for trapping become unstable and then again become stable at higher $v_0$, respectively.

In addition to explaining where the trapping probability $P_{\rm trap}$ goes to zero, certain periodic orbit bifurcations are also associated with some of the complex oscillations of $P_{\rm trap}$ when it is nonzero.
In particular, we have observed that the $t$-symmetry breaking bifurcations along the secondary branches of orbits are associated with discontinuous jumps in $P_{\rm trap}$.
This is seen for example in our calculations for $\alpha = -0.91$, shown in Fig.~\ref{fig:bifurcation_schematic}b with the $t$-symmetry breaking bifurcations marked by the yellow lines.
Here, $P_{\rm trap}$ exhibits a sizable sudden increase each time $v_0$ is increased past a $t$-symmetry breaking bifurcation.
Figure \ref{fig:tsymmbreak} shows this behavior for the range $-1 \leq \alpha \leq -0.63$, where we have detected the $t$-symmetry breaking bifurcation along the secondary branch of orbits associated with the main orbit.
In particular, we see that $P_{\rm trap}$ increases abruptly as the $t$-symmetry breaking bifurcation curve (yellow) is crossed.
We also note that $P_{\rm trap}$ abruptly decreases as $v_0$ increases past the heteroclinic bifurcation curve (green), where the attractor created in the $t$-symmetry breaking bifurcation is destroyed.
This is true even in the range of $\alpha$ where $v_0^{\rm h} < v_0^{\rm sn}$ (i.e.\ $\alpha \geq -0.67$ in Fig.~\ref{fig:tsymmbreak}).
Here, $P_{\rm trap}$ drops abruptly but remains nonzero after the heteroclinic bifurcation, and Fig.~\ref{fig:tsymmbreak} suggests that $P_{\rm trap}$ smoothly goes to zero as the saddle-node bifurcation is approached.
These observations suggest that the basin of attraction of the attractor created in the $t$-symmetry breaking bifurcation has a nonzero volume in phase space both when the attractor is created and when it is destroyed.

\section{Conclusion}\label{sec:concl}
In conclusion, we have identified the phase-space structures that cause the trapping of rigid ellipsoidal microswimmers in individual vortices of a model vortex lattice fluid flow.
At high swimming speeds $v_0$, trapping only occurs for sufficiently elongated perpendicular swimmers (i.e.\ those with $\alpha < 0$ and sufficiently close to $-1$).
Here, it is due to the asymptotic stability of certain swimming fixed points for these parameters.
These fixed points may be on the boundary of an individual vortex cell, implying that a swimmer may be localized in the vicinity of the vortex cell where it began, instead of being strictly trapped in the interior of the cell.
At low to intermediate swimming speeds, swimmers of nearly all shapes and both relative swimming directions may get trapped inside their initial vortex cell.
In phase-space, swimmers can be trapped on a quasi periodic orbit on a $t$-symmetric invariant torus surrounding a stable $t$-symmetric periodic orbit, or they can be trapped inside the basin of attraction of an asymmetric limit cycle.
We have shown numerically that the destruction (creation) of these stable solutions corresponds to the swimmer parameters  $(v_0,\alpha)$ where trapping ceases (reemerges).
In particular, we have shown that the surprising repeated breakdown and subsequent reemergence of trapping for perpendicular swimmers as $v_0$ is increased  is due to the bifurcations of certain periodic orbits and swimming fixed points.

Our investigation highlights the important role played by symmetries of the equations of motion in shaping the swimmer phase space.
In particular, reversibility (i.e.\ $t$-symmetry) plays a dominant role: it allows for the proliferation of islands of stability around stable $t$-symmetric solutions.
This is similar to the formation of KAM islands in the phase space of passive particles in 2D, time-dependent flows due to the Hamiltonian structure of the respective equations of motion.
However, reversible systems also permit dissipation, which can be triggered by $t$-symmetry breaking bifurcations that create repelling and attracting limit cycles.
We have shown that this phenomenon partially accounts for the complex oscillations of the trapping probability $P_{\rm trap}$ as the swimmer parameters are varied.
Specifically, we have provided numerical evidence that the onset of dissipation through a $t$-symmetry breaking bifurcation coincides with a discontinuous increase of $P_{\rm trap}$ as $v_0$ increases.
We have also shown that global bifurcations can occur in swimmer phase space, where periodic orbits collide with swimming fixed points.
Our numerical evidence shows that these bifurcations also explain some of the complex behavior of $P_{\rm trap}$, causing it to decrease discontinuously when attracting limit cycles are destroyed by heteroclinic bifurcations.
Of course, the oscillations of $P_{\rm trap}$ seen in Fig.~\ref{fig:monte_carlo}a have much more structure with decreasing $v_0$ that remains to be explained.
We anticipate that these oscillations occur due to a cascade of bifurcations of other periodic orbits that occur as the swimming speed decreases.

This work has focused on using the linear stability properties of solutions as indicators for the trapping or localization of swimmers around their initial vortex cell.
It would be interesting to investigate the global properties of the solutions we have identified here.
For example, how do the invariant manifolds of the unstable fixed points and periodic orbits shape the swimmer phase space?
What determines the size of the islands of stability, and are the boundaries of these islands ``sticky?"
How do these global phase-space structures---which are barriers to phase-space transport---influence the migration of swimmers between vortices?
It would also be interesting to study the interplay of rotational noise, a common feature of swimmer models, with the phase-space structures identified here.

\section*{Acknowledgments}
We thank Tom Solomon and his research group for stimulating discussions and for sharing preliminary experimental data.
We acknowledge David Brantley for his contributions during the preliminary stages of this project.
We gratefully acknowledge computing time on the Multi-Environment Computer for Exploration and Discovery (MERCED) cluster at UC Merced, which was funded by National Science Foundation Grant No. ACI-1429783.
This material is based upon work supported by the National Science Foundation under Grant No. CMMI-1825379.

\section*{Data Availability Statement}
The data that support the findings of this study are available from the corresponding author upon reasonable request.

\end{document}